\documentclass[12pt]{article}

\usepackage{scicite}
\usepackage[utf8]{inputenc}
\usepackage[T1]{fontenc}
\usepackage{graphicx}
\usepackage{subcaption}
\usepackage{lineno}
\usepackage{setspace}
\usepackage{fancyhdr}
\usepackage{times}
\usepackage{amsmath}
\usepackage{amssymb}
\usepackage{appendix}
\graphicspath{ {./images/} }

\newenvironment{sciabstract}{%
\begin{quote} \bf}
{\end{quote}}

\title{Quantum-assisted electron transport in microbial protein wires across macroscopic distances}

\author
{Jasper R. van der Veen$^{1,2\ast}$, Silvia Hidalgo Martinez$^{3}$, \\ 
Albert Wieland$^{1}$, Matteo De Pellegrin $^{1}$, Rick Verweij$^{1}$, \\
Yaroslav M. Blanter$^{1}$, Herre S.J. van der Zant$^{1}$ and Filip J.R. Meysman $^{2,3}$  \\ 
\\
\normalsize{$^{1}$Department of Quantum Nanoscience, Delft University of Technology,}\\
\normalsize{Lorentzweg 1, Delft, 2628CJ, The Netherlands}\\
\normalsize{$^{2}$Department of Biotechnology, Delft University of Technology,}\\
\normalsize{Van der Maasweg 9, Delft, 2629HZ, The Netherlands}\\
\normalsize{$^{3}$Excellence center for Microbial Systems Technology, University of Antwerp,}\\
\normalsize{Universiteitsplein 1, Wilrijk, 2610, Belgium}\\
\normalsize{$^\ast$To whom correspondence should be addressed; E-mail:  J.R.vanderVeen@tudelft.nl}
}

\date{}

\begin{document} 


\baselineskip24pt


\maketitle 

\textbf{One Sentence Summary}. Electrical conductivity in cable bacteria remains exceptionally high at cryogenic temperatures, indicative of quantum-assisted transport. 

\newpage
\begin{sciabstract}
Multicellular cable bacteria display an exceptional form of biological conduction, channeling electrical currents across centimeter distances through a regular network of protein fibers embedded in the cell envelope. The fiber conductivity is among the highest recorded for biomaterials, providing a promising outlook for new bio-electronic technologies, but the underlying mechanism of electron transport remains elusive. Here, we use detailed electrical characterization down to cryogenic temperatures, which reveals that long-range conduction in these bacterial protein wires is based on a unique type of quantum-assisted multistep hopping. The conductance near room temperature reveals thermally activated behavior, yet with a low activation energy, suggesting that substantial delocalization across charge carrier sites contributes to high conductivity. At cryogenic temperatures, the conductance becomes virtually independent of temperature, thus indicating that quantum vibrations couple to the charge transport. Our results demonstrate that quantum effects can manifest themselves in biological systems over macroscopic length scales.
\end{sciabstract}


\nolinenumbers
\section*{Introduction}
\nolinenumbers
Conventionally, the reference point for long-range biological charge transport are the protein complexes of the electron transport chain in mitochondria or the photosynthetic reaction centers in chloroplasts~\cite{moser1992nature,gray2010electron}. These membrane-bound complexes enable sequential hopping of electrons between closely spaced cofactors (e.g. hemes or iron-sulphur clusters) over total distances covering $\leq 10$~nm~\cite{blumberger2015recent}. Recent studies, however, demonstrate that biological electron transport may greatly surpass this length scale. While metal-reducing bacteria such as Geobacter and Shewanella can mediate electron transport over micrometer distances through thin surface appendages \cite{reguera2005extracellular,el2010electrical,wang2019structure}, the discovery of cable bacteria extends the range of biological electron transport to the centimeter scale ~\cite{pfeffer2012filamentous,meysman2018cable,bjerg2018long}. However, the question as to why biological protein structures can sustain conduction over these macroscopic distances remains fundamentally unresolved.

Cable bacteria harbor an internal conductive network, which consists of centimeter-long protein fibers that run in the cell envelope along the entire length of the filamentous bacteria~\cite{cornelissen2018cell,meysman2019highly,thiruvallur2020ordered,boschker2021efficient,jiang2018vitro}. Room-temperature characterisation ~\cite{meysman2019highly,bonne2020intrinsic,boschker2021efficient} reveals that these protein fibers possess an electrical conductivity up to 300~S/cm$^{-1}$,
which surpasses that of biomaterials by several orders of magnitude, and even exceeds the conductivity of most organic semiconductors \cite{schwarze2019molecular}. To elucidate the transport mechanism underlying this extraordinary form of biological long-range conduction, we characterised the fiber conductance over a wide temperature range from room temperature down to liquid helium temperature. These measurements reveal that charges are transported through a unique type of multi-step hopping, involving low barriers due to delocalized charge carrier wave functions and quantum vibration effects at temperatures below 75 K.  
\pagestyle{plain}
\nolinenumbers
\section*{Temperature dependence of conductance}
\nolinenumbers

\begin{figure}[t!]
\centering
\includegraphics[width=0.8\linewidth]{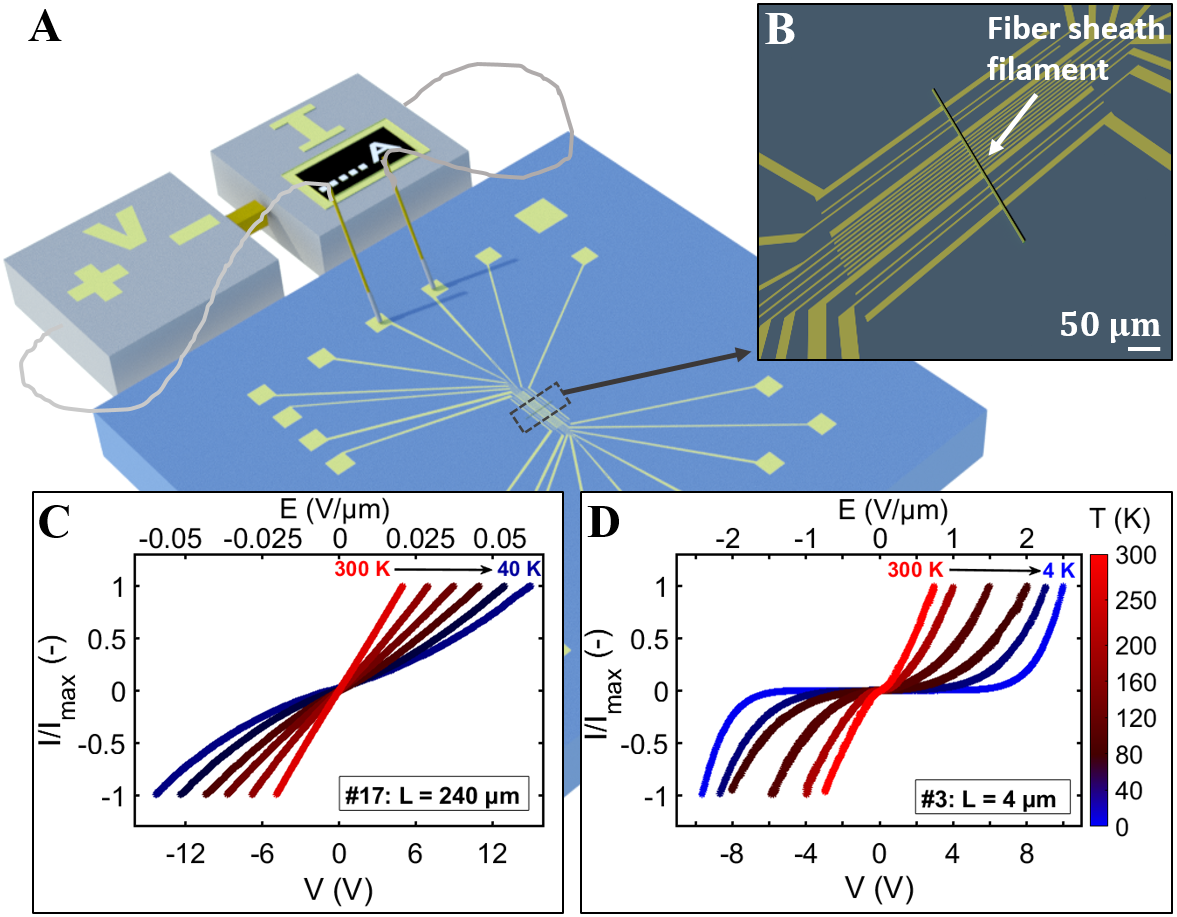}
\caption{\textbf{Current as a function of temperature and electric field.} (\textbf{A}) Individual fiber sheaths are deposited on Si/SiO$_2$ substrate with prepatterned gold contacts. (\textbf{B}) A fiber sheath filament is stretched across a series of gold contacts. (\textbf{C} and \textbf{D}) Representative current ($I$) versus bias voltage ($V$) curves at different temperatures for two different segment lengths, $L$. The  current is normalized to the maximum current of each $I(V)$ trace. Temperature is indicated by the color scale. The top axis displays the applied electric field, $E=V/L$. Shorter segments allow to investigate the conductance at lower temperatures and higher electrical fields. For the longer segment ($L =240$~µm), the current fell below the detection limit below 40~K.}
\label{fig:2}
\end{figure}

Conductivity probing as a function of temperature is widely applied to elucidate the charge transport mechanism in electronic materials \cite{Liu2017AUU}. For biological materials, investigation of low-temperature conduction ($<10$ K) is not physiologically relevant, but is instrumental to get insight into the mechanism of electron transport in proteins, as thermal excitation becomes suppressed. Up until now, cryogenic characterization of electron transfer has been primarily applied to photosynthetic reaction centers~\cite{de1966studies,devault1967electron}, which are experimentally accessible through facile light activation and high electron transfer rates. For other protein systems, however, data on low-temperature conductance are critically missing, and consequently, it remains uncertain to what extent insights and parameters from photosynthetic reaction centers can be extrapolated \cite{moser2010guidelines}. In particular, low-temperature conduction studies are lacking for macroscale ($\geq 1~\mu$m) protein structures, as the increasing resistance with length prevents current detection at low temperatures. The high conductance sustained across millimetres in cable bacteria alleviates this experimental bottleneck.

To probe the conductance of the periplasmic fibers, long cable bacterium filaments ($>2$~mm) were individually isolated from enrichment cultures. Subsequent extraction provides a so-called "fiber sheath" that retains the conductive fiber network lying on top of a connective carbohydrate sheath.\cite{meysman2019highly} The number of parallel fibers embedded in one fiber sheath is determined by microscopy (fig.~S\ref{fig:EXT_SEM}). Combined with the known~\cite{boschker2021efficient} fiber diameter (26~nm), this enables the fiber conductivity to be calculated from the measured conductance (Eq.~\ref{eq:conductivity} in Supplementary Text). This procedure has been shown to provide reliable fiber conductivity data \cite{meysman2019highly,bonne2020intrinsic}, as native filaments and fibers sheaths provide similar fiber conductivities, indicating that the extraction procedure does not affect the electrical response \cite{meysman2019highly}.       

Fiber sheaths of variable length (diameter 4~µm; length up to several mm) were deposited onto insulating silicon substrates with pre-patterned gold electrodes (Fig.~\ref{fig:2}A) for electrical characterization under vacuum. The multiple electrical contacts per filament allow conduction to be evaluated over different segments of the fiber sheath (Fig.~\ref{fig:2}B). Room-temperature, four-probe measurements on single filaments show a linear dependence of the resistance $R$ on the probed segment length $L$ over a distance of millimeters (fig.~S\ref{fig:EXT_FIG_LengthDep}). These measurements demonstrate that the fiber network possesses a uniform conductivity over mm-scale distances, which hence justifies the comparison of different segments.

Current-voltage characteristics ($I(V)$ curves) were recorded for 53 segments of varying length ($L$ = 4~µm to 2~mm), of which 16 were also measured down to cryogenic temperatures ($4.2-10$~K). Measurements were highly consistent between filaments, revealing a similar response. Figure~\ref{fig:2}B and D displays representative $I(V)$ data for both a long ($L=240$~µm) and short segment ($L= 4$~µm) (fig.~S\ref{fig:IV_GT_All} provides data for all segments). The measured current shows a distinct response to the temperature, $T$, and to the imposed electric field, $E=V/L$. At high temperature and low electric field, the $I(V)$ curve is linear (Fig.~\ref{fig:2}B). Yet, when temperature decreases and the electric field strength increases, the $I(V)$ becomes increasingly more non-linear (Fig.~\ref{fig:2}C and D). 

In a two-probe configuration, the measured resistance accounts for the intrinsic resistance of the probed filament segment as well as the contact resistance between the gold electrodes and filaments\cite{meysman2019highly}. To verify our approach, two-probe and four-probe measurements were compared on the same segment, revealing that contact resistances do not influence the shape of the $I(V)$ profile, nor the temperature dependence of the conductance (fig.~S\ref{fig:HighT_G_Analysis}). 

\begin{figure}[h!]
    \begin{subfigure}[t]{0.49\textwidth}
        \raggedleft
        \includegraphics[scale=0.37]{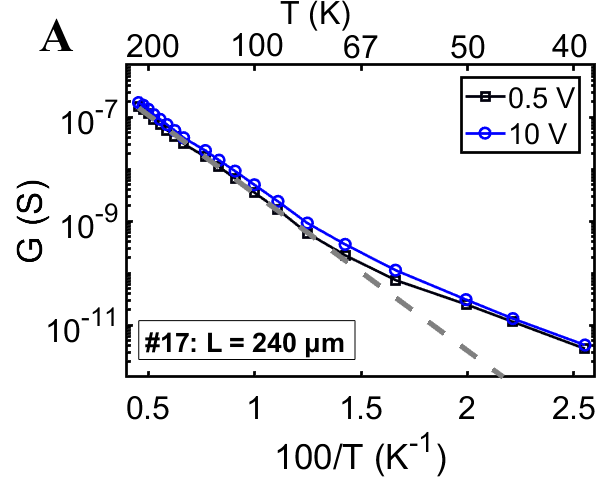}
    \end{subfigure}
    \hspace{0.5cm}
    \begin{subfigure}[t]{0.49\textwidth}
        \raggedright
        \includegraphics[scale=0.37]{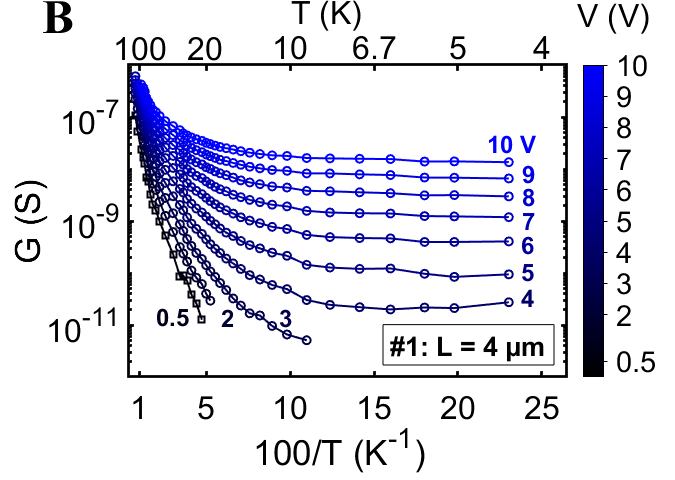}
    \end{subfigure}
    
    \vspace{0.25cm}
    \begin{subfigure}[t]{0.49\textwidth}
        \raggedleft
        \includegraphics[scale=0.37]{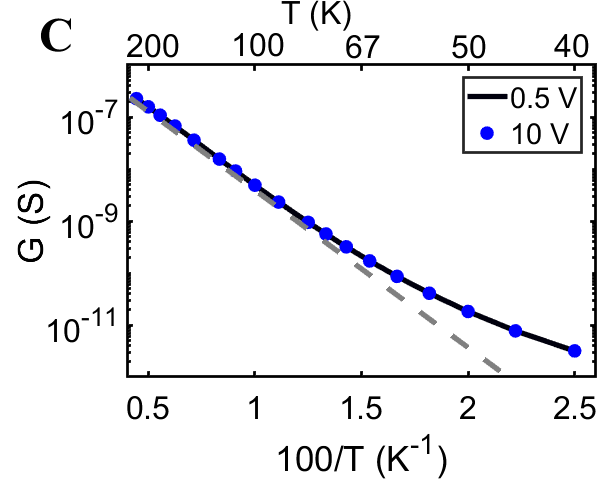}
    \end{subfigure}
    \hspace{0.5cm}
    \begin{subfigure}[t]{0.49\textwidth}
        \raggedright
        \includegraphics[scale=0.37]{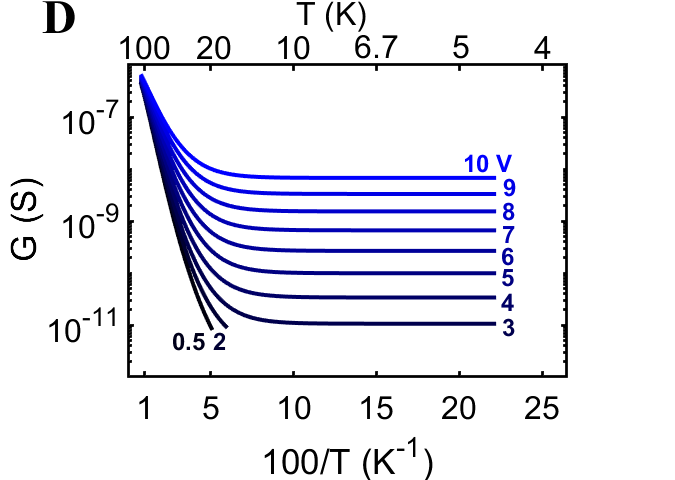}
    \end{subfigure}

    \begin{subfigure}[t]{0.49\textwidth}
        \raggedleft
        \includegraphics[scale=0.37]{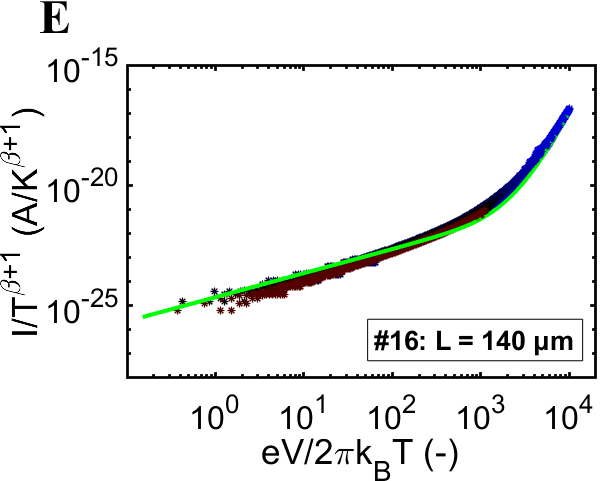}
    \end{subfigure}
    \hspace{0.5cm}
    \begin{subfigure}[t]{0.49\textwidth}
        \raggedright
        \includegraphics[scale=0.37]{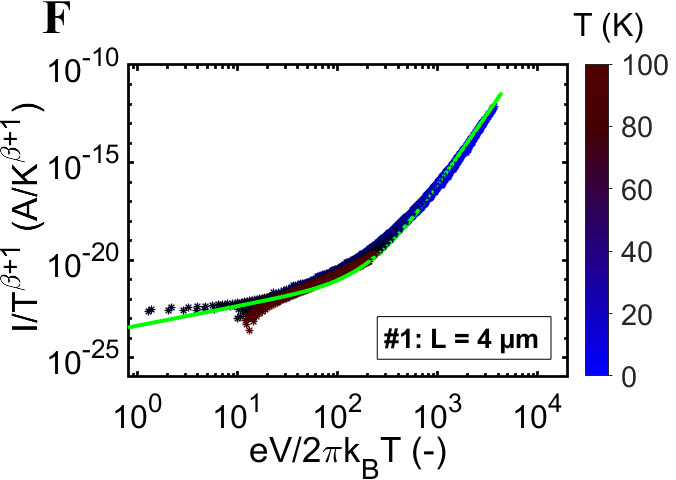}
    \end{subfigure} 
    
\caption{\textbf{Dependence of conductance on temperature and electrical field.} (\textbf{A} and \textbf{B}) Conductance data $G=I/V$ are plotted as a function of inverse temperature at a constant electric field strength (i.e. fixed bias voltage) for two segments of different length. Data are extracted from $I(V,T)$ curves as shown in Fig. 1. In panel (A), the Arrhenius fit to the high-temperature conductance data is included for reference (dashed grey line). (\textbf{C} and \textbf{D}) Model simulation of the conductance data in panels (A) and (B) via a 1D hopping chain model with one effective vibrational mode (Jortner model). Model parameters in panel (C): reorganization energy, $\lambda = 0.35$ eV and effective vibrational frequency, $\langle\omega\rangle =$ 122 cm$^{-1}$ (15 meV). The number of hopping sites, $N_{\rm S}$, does not change the shape of the curve. Model parameters in panel (D):  $\langle\omega\rangle =$ 58 cm$^{-1}$ (7.2 meV), $\lambda = 0.16$ eV, number of hopping sites, $N_{\rm S} = 200$. (\textbf{E} and \textbf{F}) $I(V,T)$ data replotted as a universal scaling curve \cite{egger1994quantum,Asadi2013PolaronHM}; $T$ is temperature,  $e$ is the elementary charge, and $k_{\rm B}$ is Boltzmann's constant. The green solid line represents a model fit to the data that has two fitting parameters: the dimensionless exponent $\beta$ and the number of hopping sites, $N_{\rm S}$. For the longer segment, $N_{\rm S} =1400$ and $\beta =6.0$; for the shorter segment, $N_{\rm S} = 120$ and $\beta = 6.5$.}
\label{fig:3}
\end{figure}

When replotting the $I(V)$ data in terms of the conductance, $G=I/V$, we consistently observe the same $G(T,E)$ response across all segments (Fig.~\ref{fig:3}A and B; fig.~S\ref{fig:IV_GT_All}). At high temperatures ($T>75~$ K), the $G(T)$ data collected at different $E$-values converge  (Fig.~\ref{fig:3}B), so conductance is not affected by the electric field. Moreover, the charge transport is thermally activated and follows a classical exponential Arrhenius-type dependence~\cite{bonne2020intrinsic}. Below the cross-over temperature $T_{\rm C} \backsim 75~K$, the temperature dependence becomes noticeably weaker, and for T < 20 K, the conductance remains high and becomes virtually independent of temperature at moderate electric fields (Fig.~\ref{fig:3}B). At these low temperatures, the conductance is strongly affected by the electric field. For example, at 5~K, $G$ increases by 3 orders of magnitude when $E$ is increased from 1 to 2.5~V/µm (Fig.~\ref{fig:3}B).

\clearpage

\nolinenumbers
\section*{Quantum vibrations couple to the electron transport}
\nolinenumbers

The observation that the conductance remains elevated as the temperature decreases, and eventually becomes independent of temperature, is remarkable.  The deviation from an exponential temperature dependence is at odds with the standard semi-classical Marcus theory of electron transfer~\cite{Marcus1985ElectronTI}, which assumes that transitions are thermally activated. The fact that the conductance remains high at low temperatures thus hints at an energy source other than thermal energy that assists the charge transfer. In seminal experiments on photosynthetic reaction centers in purple bacteria, DeVault and Chance observed a similar temperature dependence that could not be accounted for by Marcus theory. The charge separation reactions of photosynthesis continued at cryogenic temperature and the rate of these reactions became independent of temperature~\cite{de1966studies,devault1967electron}. 

This finding spurred a host of theoretical developments, which eventually extended electron transfer theory to a fully quantum-mechanical treatment~\cite{hopfield1974electron,Jortner1976TemperatureDA,egger1994quantum,van2011quantum}. Quantum theory asserts that at absolute zero, molecular vibrations retain a finite zero-point energy. At low temperatures, thermal motion is frozen out, but quantised vibrational modes can assist the electron tunnelling, in a process referred to as nuclear tunneling~\cite{hopfield1974electron,Jortner1976TemperatureDA}. Marcus theory\cite{Marcus1985ElectronTI} accounts for electron tunneling, but uses a classical harmonic approximation of the nuclear motion, so it neglects the impact of nuclear tunneling and therefore underestimates the electron transfer rate at low temperatures.

To verify whether nuclear tunnelling could explain our observations, we modelled electron transport subject to quantum vibrational coupling in a one-dimensional hopping chain that mimics the conduction path in the fiber sheaths (Supplementary Text). In metalloproteins, vibrational coupling can involve both high-frequency intramolecular vibrations of the charge carrying cofactor as well as lower-frequency vibrations of the surrounding protein matrix and solvent molecules. This vibrational coupling can be described with models of increasing complexity. As a first step, we applied a model where only one effective high-frequency mode $\langle\omega\rangle$ stimulates the hopping process (Jortner model~\cite{Jortner1976TemperatureDA}, Supplementary Text). These model calculations adequately capture the essence of our experimental data: the conductance displays two regimes with a transition at the cross-over temperature $T_{\rm C} = \hbar \langle\omega\rangle /k_{\rm B}$, where $k_{\rm B}$ is the Boltzmann constant and $\hbar$ the reduced Planck constant. For $T>T_{\rm C}$, the vibrational mode remains thermally excited, and the conductance adopts a classical Arrhenius-type dependence (Fig.~\ref{fig:3}C). Below $T_{\rm C}$, the vibrational mode is no longer thermally excited, but due to the quantum vibrational energy, the conductance remains higher than predicted by the Arrhenius relation. Upon further cooling below $T_{\rm C}/5 \approx$ 25 K, the conductance becomes virtually independent of temperature (Fig.~\ref{fig:3}D).   

Across all sixteen segments investigated, we find a mean cross-over temperature $T_C$ = 75 $\pm$ 14 K, which corresponds to a  characteristic frequency $\langle \omega \rangle= k_{\rm B} T_C/\hbar$ = 52 $\pm$ 10  cm$^{-1}$ for the single effective vibrational mode that couples to the electron hopping (vibrational energy $\hbar\langle \omega \rangle$  = 6.4 $\pm$ 1.2 meV). The experimental verification of vibronic coupling is highly challenging in biological systems, and has only been done for photosynthetic reaction centers, which yields substantially higher characteristic frequencies $\hbar \langle \omega \rangle \approx 60$ meV~\cite{jortner1980dynamics}. Previously, it has been speculated that the $60$ meV frequency could be a general feature of electron tunneling reactions in natural proteins~\cite{moser2010guidelines}. Our results invalidate this hypothesis, and instead, suggest a substantial variability in vibrational coupling, where the characteristic frequency is likely dependent on the specific molecular configuration of the cofactor and metalloprotein involved in the electron transport. To illustrate this, fig.~S\ref{fig:ChromatiumCBJortner} compares the data fit for a cable bacterium to that of the photosynthetic reaction center in \emph{Allochromatium vinosum}.

\begin{figure}[h!]
\centering
    \begin{subfigure}[t]{0.45\textwidth}
        \includegraphics[scale=0.37]{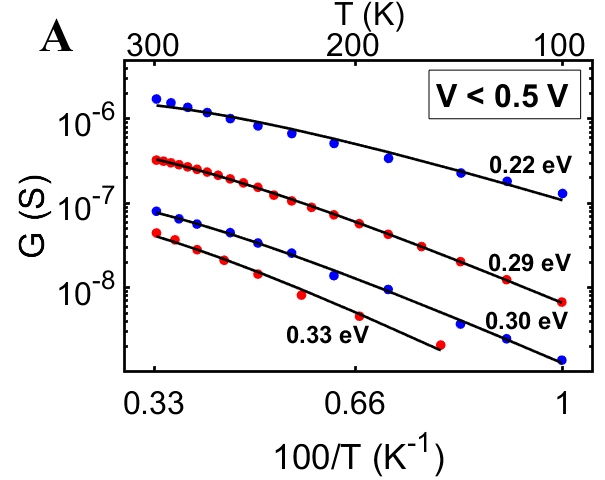}
        \label{fig:MarcusFits}
    \end{subfigure}
    \begin{subfigure}[t]{0.45\textwidth}
        \includegraphics[scale=0.37]{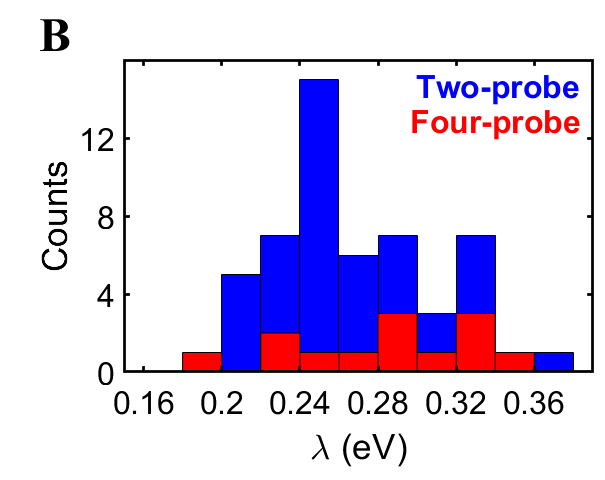}
    \end{subfigure}
    \begin{subfigure}[t]{0.45\textwidth}
        \includegraphics[scale=0.37]{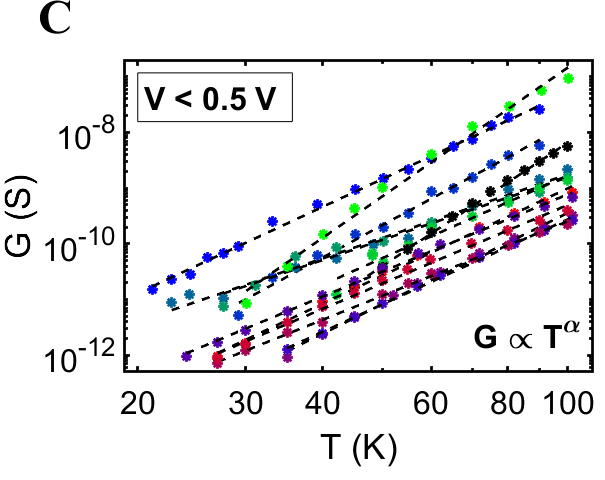}
    \end{subfigure}
    \begin{subfigure}[t]{0.45\textwidth}
        \includegraphics[scale=0.37]{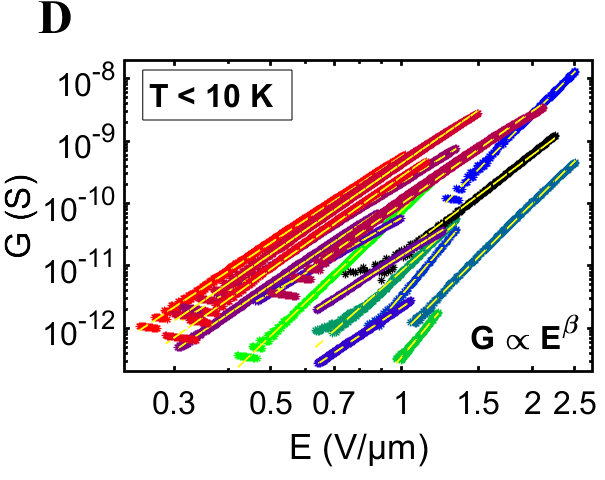}
    \end{subfigure}    
\caption{\textbf{Empirical fits to conductance data.} (\textbf{A}) Low-bias conductance, $G$, as a function of inverse temperature for $T > 100$~K. Data for 4 segments. Two- and four-probe measurements are shown in blue and red markers respectively.  Solid black lines are fits by the non-linear least squares method to the Marcus equation $G=g_0 (k_{\rm B} T)^{-3/2}\text{exp}(\frac{\lambda}{4k_{\rm B}T})$, where $g_0$ is a pre-factor and $\lambda$ the reorganization energy, of which the value is given next to the data. (\textbf{B}) Histogram with reorganization energies for $n=53$ samples (blue: two-probe measurements; red: four-probe measurements). (\textbf{C}) Low-bias conductance against temperature in the region $20-100$~K for 16 segments. Black dashed lines represent power law fits  (determining  $\alpha$). (\textbf{D}) Conductance versus applied electric field at the lowest temperature achieved ($T < 10$~K), for the same 16 segments. Yellow dashed lines represent power law fits (determining $\beta$).}
\label{fig:4}
\end{figure}
 
\nolinenumbers
\section*{Conductance follows a universal scaling relation}
\nolinenumbers

The single-mode vibrational model predicts that the zero-bias conductance should become constant at the lowest temperatures. Yet closer data inspection reveals that the conductance continues to decrease as a power law ($G\propto T^\alpha$; Fig.~\ref{fig:4}C), which suggests that vibrational modes lower than $\omega = k_{\rm B} T_C/\hbar$ are additionally coupling to the electron transport. To accommodate for this,  we integrated a multi-mode description of vibrational coupling in our one-dimensional hopping model (Egger model \cite{egger1994quantum}, Supplementary Text). There are several predictions that can be tested. At high temperatures, all vibrational modes must remain thermally excited, and hence, one should retrieve the results of Marcus theory. This is indeed the case. For $T> 100$ K, the conductance follows a Marcus-type dependence on temperature, $G \backsim  T^{-3/2}\text{exp}(-\lambda/4k_{\rm B}T)$, where $\lambda$ is the reorganization energy (Fig.~\ref{fig:4}A; Supplementary Text). Also, at high temperatures, the conductance should remain independent of $E$, which provides linear $I(V)$ curves, as observed (Fig.~\ref{fig:2}B). The reorganization energy is low and shows limited variation (Fig.~\ref{fig:4}B), $\lambda = 270 \pm 40$~meV ($n = 53$ segments). The corresponding activation energy, $U_{\rm A} =\lambda/4 = 42 \pm 8$~meV (fig.~S\ref{fig:Arrhenius}), aligns well with earlier high-temperature electrical measurements on both intact cable bacteria and fiber sheaths~\cite{bonne2020intrinsic}.       

The multi-mode model predicts that the conductance should follow a power-law dependence on temperature at low electric fields, $G \propto T^{\alpha}$, and a power-law dependence on the electric field at low temperatures, $G \propto E^{\beta}$. The fiber network conductance indeed shows such a power law behaviour with exponents $\alpha \approx 3.7 - 7.7$ and $\beta \approx 4.0 - 8.3$ (Fig.~\ref{fig:4}C and D; table~S\ref{table:NT}). Moreover, the multi-mode model asserts that in the low-temperature regime, the $I(V,T)$ data for a given segment can be suitably reduced to a single universal scaling curve~\cite{Asadi2013PolaronHM} (see Supplementary Text), which features the normalized current $I/T^{1+\beta}$ as a function of a normalized bias $\overline{V} ={eV}/{(2\pi k_{\rm B}T)}$. The rescaled I/V data closely follow this universal scaling curve for all 16 segments investigated (Fig.~\ref{fig:3}E and F; fig.~S\ref{fig:USC_All}). This universal scaling behavior has not been seen before for a biological material, but is characteristic for various synthetic one-dimensional conductors, such as networks of graphene nanoribbons\cite{richter2020charge} and carbonized polymer nanofibers\cite{kim2016apparent}, which display similarly high conductivities as the periplasmic fibers in cable bacteria. Moreover, the power law exponent $\beta$, which designates the degree of vibronic coupling, is on par with these synthetic conductors, thus suggesting that cable bacteria may have evolved a similar electron relay system.    

\nolinenumbers
\section*{A unique form of multistep hopping}
\nolinenumbers

Our results provide several insights into the mechanism that sustains long-range conduction in the periplasmic fiber network of cable bacteria. Foremost, the observed temperature dependence of conduction displays the hallmark of quantum-mechanical electron tunneling \cite{moser2010guidelines}. At high temperatures, electron transport is assisted by thermal fluctuations, thus providing an Arrhenius dependence, while at low temperatures, quantized molecular vibrations couple to the electron transfer, thus resulting in higher conductance than classically expected. Nuclear tunneling has been previously observed in photosynthetic reactions centers \cite{devault1967electron}, but not in macroscale electron transfer. 

Electron tunneling shows a strong exponential dependence of the electron transfer rate on distance, which limits individual electron transfer events to distances $\le$ 1.5 nm \cite{moser2010guidelines}. Still, it is well known that protein structures can support electron transport over much longer distances, of which the cm-scale conduction in cable bacteria represents the most extreme example. Biology has resolved this problem by arranging cofactors in chains at close spacings (typically $10 - 15$ Å). This enables multi-step electron hopping, in which electrons are moving through the protein matrix by consecutive tunneling steps, as for example, documented for multiheme cytochromes involved in extracellular respiration~\cite{blumberger2015recent} and FeS clusters aligned in the complexes of the respiratory chain \cite{read2021mitochondrial}. While our data support this multi-step hopping picture, there are several unique features in which the electron transport in cable bacteria is distinct from the currently known forms of long-range electron transport in proteins. 

Foremost, the length scale over which hopping takes place in cable bacteria is extraordinary, as it covers distances of millimetres to centimeters. Raman spectroscopy has shown that the periplasmic fiber network does not contain FeS clusters nor cytochromes \cite{meysman2019highly,boschker2021efficient}, thus excluding a heme-based conduction mechanism as found in the surface appendages of metal-reducing bacteria \cite{filman2019cryo,pirbadian2014shewanella,wang2019structure}. Instead, recent work suggests that the periplasmic fibers in cable bacteria contain a nickel-sulfur cofactor that mediates the electron transport \cite{boschker2021efficient}. The involvement of nickel is remarkable, as all currently known metalloproteins involved in electron transport rely on either iron- or copper-containing cofactors, but never nickel \cite{liu2014metalloproteins}. It is thus reasonable to suggest a model in which the quantized vibrational modes of this nickel-sulfur cofactor are strongly coupled to the electron transport. If such vibronic coupling is due to a single mode, this mode would have characteristic frequency of $\omega \approx$ 50 cm$^{-1}$. The corresponding Huang-Rhys factor, which provides a dimensionless measure for the coupling strength, is high ($S=\lambda / \hbar \langle \omega \rangle$ = 42), thus indicating strong vibronic coupling. While the molecular structure of nickel cofactor remains unresolved, the appearance of distinct Ni-associated Raman peaks in the spectra of fiber sheaths near this frequency \cite{boschker2021efficient} suggests that cofactor modes could be indeed involved in the reorganization process \cite{myers1996resonance}. Most likely, multiple modes are coupling to the electron transport, which then provides the observed universal scaling response (Fig.~\ref{fig:3}E and F) as also seen in abiotic one-dimensional conductors \cite{Asadi2013PolaronHM,choi2012probing,kim2016apparent}.  

A second notable feature is that the reorganization energy at room temperature (0.27~eV) is markedly lower than observed for other forms of biological electron transport. Typical values for single electron transfer steps in enzymes are in the range 0.7 < $\lambda$ < 1.4~eV, while multi-step electron transport in the surface appendages of the metal-reducing bacteria also exhibits similarly high reorganization energies~\cite{xu2018multiheme,malvankar2011tunable}. Within the thermally activated regime, and assuming a negligible driving force (as the voltage bias is distributed over very many transitions in the hopping chain), the rate of electron transfer scales with $\sim \exp(\lambda/4 k_{\rm B} T)$, and so, a low reorganization energy helps explaining the extraordinary high fiber conductivity observed in cable bacteria. Moreover, the reorganization energy is known to inversely scale with the localisation length of the charge carriers~\cite{taylor2018generalised}. Therefore, high reorganization energies imply that electrons are strongly localized on the charge carrier site, while low reorganization energies, as seen here, suggest substantial delocalization of the charge carrier wave function. Heme-localised electronic states are examples of the former: they form low energy states (“traps”) that are separated by sizable activation barriers of 0.2 $-$0.3 eV, which are large compared to thermal activation barrier ($\approx$ 10 $k_{\rm B} T$ at 300 K) \cite{blumberger2015recent}. In contrast, the electronic states on the cable bacterium fibers appear to be far more delocalized, thus enabling fast transitions. Comparably low reorganization energies have been reported for highly-mobility organic semiconductors, such as rubrene and naphtalene\cite{giannini2019quantum}, in which the charge carrier sites display sizeable conjugation and delocalization. 

\nolinenumbers
\section*{The challenge of high conductivity}
\nolinenumbers

Overall, the observed temperature dependence is consistent with  multistep hopping, which forms the cornerstone of the existing theories of long-range biological electron transfer \cite{blumberger2015recent}. Still, the high fiber conductivity recorded (up to $\backsim$ 100 S cm$^{-1}$; fig.~s\ref{fig:EXT_FIG_LengthDep}) poses a profound challenge to this hopping model. The key problem is the magnitude of the electron transfer rate $\Gamma_R$, which is related to the fiber conductivity $\sigma$ via the relation 
\begin{linenomath}
\begin{equation}
\Gamma_{\rm R} = \sigma \frac{k_{\rm B} T}{e^2} \frac{\pi N_{\rm S} d_{\rm F}^2}{L N_{\rm C}},
\end{equation}
\end{linenomath}
In nearly all electron transfer reactions in biology, the center-to-center spacing between charge carriers falls within the range of $a = 0.5-2$ nm. Consequently, one needs $N_{\rm S} =$~5-20 x $10^{6}$ sequential charge carriers to cover the typical $L$ = 1 cm length scale of a cable bacterium filament. Even if we assume massive parallelization of conduction channels (each fiber consists of $N_{\rm C} =$~125 parallel channels of 2 nm diameter; Supplementary Text), the required transition frequency for nearest neighbor hopping exceeds  10$^{13}$ s$^{-1}$ (see details in Supplementary Text). This < 0.1 picosecond electron transfer rate is orders of magnitude faster than currently observed for non-light driven electron transfer, which maximally reaches nanoseconds for inter-heme transfer in cytochrome c oxidase, but is more typically on the scale of microseconds  \cite{blumberger2015recent}. Effectively, the 10$^{13}$ s$^{-1}$ hopping rate exceeds the speed limit of non-adiabatic electron transfer, where relaxation times of vibrational modes are slower than the hopping rate itself \cite{polizzi2012physical,moser2010guidelines}.
  
Our data hint at a possible resolution to this problem. A larger center-to-center distance would allow for the same conductivity with a smaller hopping rate, thus bringing the electron transport back in the non-adiatabic regime. The universal scaling curve displays a cross-over point between high and low temperature regimes, which enables a direct estimate for the number of hopping sites ($N_{\rm S} = eV/(2\pi k_{\rm B} T)$). We find that $N_{\rm S}$ linearly scales with the segment length $L$ (table~S \ref{table:NT}; fig.~S\ref{fig:NL}), thus providing a center-to-center distance between hopping sites in excess of 10~nm. Clearly, this distance is far too large to enable through-space tunneling of electrons \cite{binnig1982vacuum,frisenda2018quantum}, and largely exceeds the known heme-to-heme distances in cytochromes (< 1 nm) \cite{mowat2005multi,breuer2014electron}. Still, in doped organic semiconducting nanowires, like polyacetylene, the universal scaling curves provides similar center-to-center distances ($\sim 10$ nm) \cite{Asadi2013PolaronHM,choi2012probing,kim2016apparent}. A potential model that can accommodate these large center-to-center distance consists of hopping chains embedding stacks of tightly spaced cofactors (> 10 nm). Electron tunneling then occurs across low-potential barriers between the cofactor stacks, while in the internal part of the stacks, electrons are delocalized and electron transport occurs largely unobstructed (fig.~S\ref{fig:HoppingSchematic}). This speculative model provides a potential explanation of how the high electrical conductivity in the periplasmic fibers in cable bacteria can be reconciled with the multistepping mechanism that emerges from data obtained here, but requires further experimental validation. Ultimately, such a better understanding of the charge carrier transport in cable bacteria may enable the design and construction of new bio-mimetic materials for electronics and energy conversion~\cite{jortner1997molecular,blankenship2011comparing,balzani2001electron}.

\newpage
\nolinenumbers
\bibliography{scibib}
\bibliographystyle{Science}

\newpage
\nolinenumbers
\section*{Acknowledgements}
\nolinenumbers
We thank Spiros Skourtis for discussions. 

\nolinenumbers
\subsection*{Funding}
\nolinenumbers
J.R.V. and F.J.R.M. were financially supported by the Netherlands Organization for Scientific Research (VICI grant 016.VICI.170.072) F.J.R.M. and S.H.M. were additionally supported by Science Foundation Flanders (FWO grant G043119N). F.J.R.M. and H.S.J.Z. received support from the EIC Pathfinder project PRINGLE.

\nolinenumbers
\subsection*{Author contributions}
\nolinenumbers
S.H.M. performed cable bacteria culturing and fiber sheath preparation for all experiments. J.R.V., A.W., M.P. and R.V. carried out conductance experiments. J.R.V. analysed conductance data and performed modelling, while H.S.J.Z., F.J.R.M., and Y.M.B. provided additional input. J.R.V., H.S.J.Z., and F.J.R.M. wrote the manuscript with additional contributions of all other co-authors.

\nolinenumbers
\subsection*{Competing interest}
\nolinenumbers
The authors declare no competing interests.

\nolinenumbers
\subsection*{Data and materials availability}
\nolinenumbers
All data are made available. The complete data set for each figure is provided with an accompanying explanation file.

\newpage
\setcounter{page}{1}
\resetlinenumber[1]
\nolinenumbers
\section*{Supplementary Materials}
\nolinenumbers
Materials and Methods \\
Figs. S1 to S9 \\
Tables S1 and S2 \\
Supplementary Text \\
\nolinenumbers

\newpage 
\nolinenumbers
\section*{Materials and Methods}

\subsection*{Conductive fiber networks from cable bacteria}
\nolinenumbers
Cable bacterium filaments were harvested from enrichment cultures that were set up using natural sediment collected in the creek bed of a salt marsh (Rattekaai, The Netherlands). Upon collection, sediment was sieved and repacked into PVC core liner tubes (diameter 40 mm). The cores were incubated in aerated artificial seawater at \emph{in situ} salinity, and the development of cable bacteria was tracked by microsensor profiling and microscopy (procedure as in Ref. 11). Under a stereo microscope, individual filaments were gently pulled out from the top layer of the sediment with custom-made glass hooks.   

Through sequential extraction, the conductive fiber network is isolated from the cell envelope of individual filaments. This extraction removes the membranes and cytoplasm, but retains the parallel conductive fibers embedded in a basal sheath (procedure as in Ref. 8). These so-called "fiber sheaths" form the starting material for all investigations performed here. To produce these fiber sheaths, freshly isolated cable bacterium filaments were cleaned by transferring them at least six times between droplets ($\backsim$ 20 µl) of MilliQ water on a microscope cover slip. Subsequently, filaments were extracted in a droplet of 1 (w/w) sodium dodecyl sulfate (SDS) for 10 minutes, followed by six MilliQ washes. Filaments were then incubated for 10 minutes in a droplet of 1 mM sodium ethylene diamine tetra-acetate (EDTA), pH 8, and again six times washed in MilliQ. 

\nolinenumbers
\subsection*{Electrical characterization down to cryogenic temperatures}
\nolinenumbers
Gold electrode patterns (Fig.~\ref{fig:2}) were deposited onto \text{$p^{++}$}-doped silicon substrates with a surface layer of silicon dioxide (285 or 500~nm thickness) via optical lithography. A laser writer illuminates the desired pattern in a single light-sensitive resist layer (AZ ECI 3007 or 3012). The laser wave length (365~nm) limits the minimum feature size to approximately 1 µm. The gold thickness was 100~nm, with 5 nm of titanium underneath to promote adhesion of the gold layer to the SiO$_2$ surface. Fiber sheaths were positioned onto patterned substrates immediately after extraction and substrates were directly transferred to the vacuum chamber of the probe station. In this way, fiber sheaths retain a stable conductance for up to a period of weeks \cite{meysman2019highly}. 

The conductance of fiber sheaths was measured down to liquid helium temperatures in two separate set-ups. The first setup is based around a dewar of helium, in which a stick containing a vacuum sample chamber can be inserted. In this setup, the patterned substrates were glued to a chip carrier, either with silver paint or epoxy glue. The electrode pads of the substrate were subsequently wire bonded to the electrodes of the chip carrier.
For the lowest temperatures ($T$ < 20 K) the sample chamber is not completely vacuum, because helium exchange gas is used to reach the desired temperature. The minimum temperature reached is near the helium condensation point (4.2 K). For higher temperatures ($T$ > 20 K), this exchange gas was pumped out of the sample chamber. A resistor was used to heat up the device, while a thermometer is placed nearby the sample to measure the temperature. The second set-up used was a cryo-free LakeShore Cryogenic Probe Station (Type CRX 6.5 K). In this case, the electrode pads did not need to be wire bonded. Of the 16 segments for which the temperature dependence was measured down to the lowest temperatures ($T$ < 10 K; fig.~S\ref{fig:IV_GT_All}), segments 1 - 5 were measured in the first set-up and segments 6 - 16 were measured in the second.

To assess the impact of contact resistances, two-probe and four-probe measurements were conducted. In a two-probe measurement, a bias voltage, $V_{\rm B}$, is applied across two electrodes, and the induced current, $I_{\rm M}$, is measured. This yields the two-probe resistance, $R_{2P} = V_{\rm B}/I_{\rm M} =  R_{\rm i} + R_{\rm C}$, which is composed the intrinsic resistance of the fiber sheath segment, $R_{\rm i}$, and the two contact resistances between the fiber sheath and the electrodes, $R_{\rm C}$. In the four-probe approach, the bias current, $I_{\rm B}$, is injected over the two outer electrodes, while the voltage, $V_{\rm M}$, is measured over the two inner electrodes. The four-probe resistance, $R_{\rm 4P}$ = $V_{\rm M}/I_{\rm B}$, equals the intrinsic resistance of the conductive fiber sheath segment between the two inner pads. The contact resistance is hence determined as $R_{\rm C} = R_{\rm 2P}-R_{\rm 4P}$.

\begin{appendices}
\setcounter{figure}{0}   

\newpage
\nolinenumbers

\section*{Figs. S1 to S9}

\subsection*{Figure S1}
\begin{figure}[h!]
\centering
\includegraphics[width=0.99\linewidth]{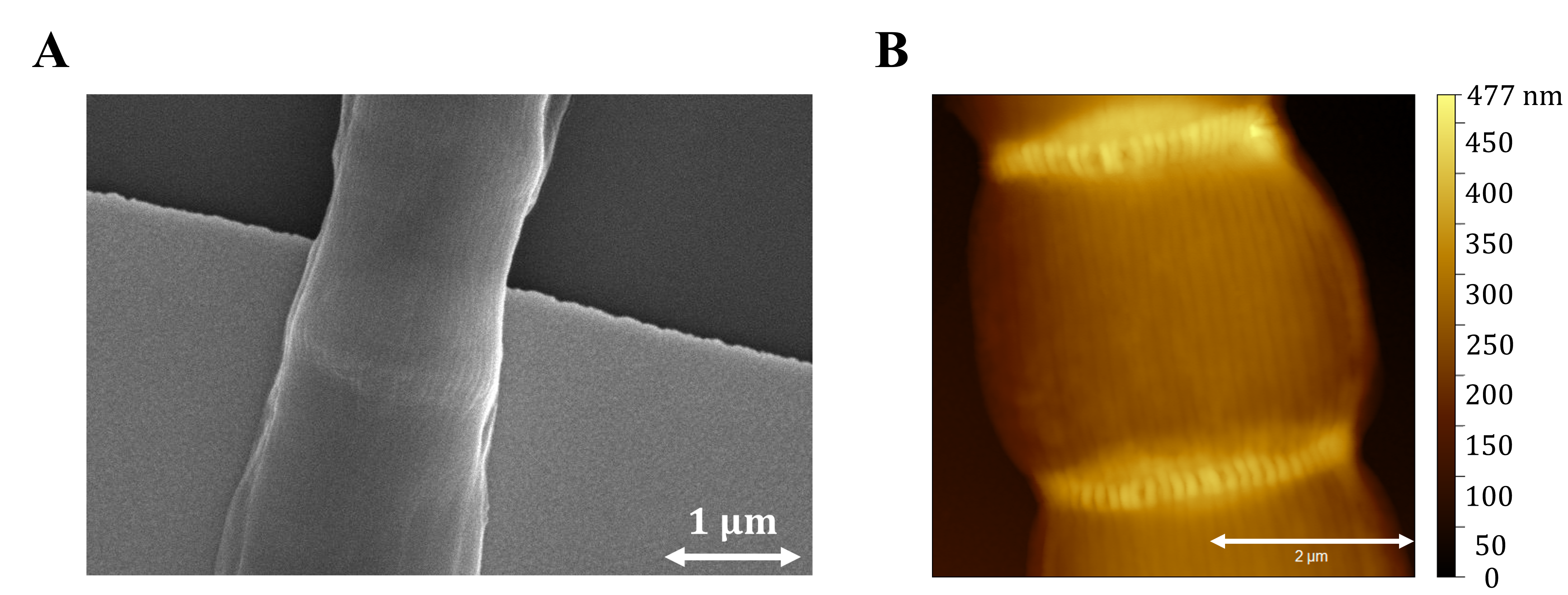}
\caption{\textbf{Microscopic images of fiber sheaths from cable bacteria}. (\textbf{A}) Scanning electron micrograph (acceleration voltage 10~kV) of a fiber sheath deposited silicon-dioxide substrate with gold patterned electrodes. The scale bar is indicated in the bottom right corner of the image. The dark background is silicon dioxide, the lighter grey background is the gold electrode.  (\textbf{B}) Atomic force microscopy image of a fiber sheath filament. The colour indicates the measured height profile (see colour bar). The fiber sheath displays the conductive fibers as a set of lines running in parallel to the bacterial filament.} 
\label{fig:EXT_SEM}
\end{figure}

\newpage
\subsection*{Figure S2}
\begin{figure}[h]
    \centering
    \begin{subfigure}[t]{0.45\textwidth}
        \raggedleft
        \includegraphics[scale=0.37]{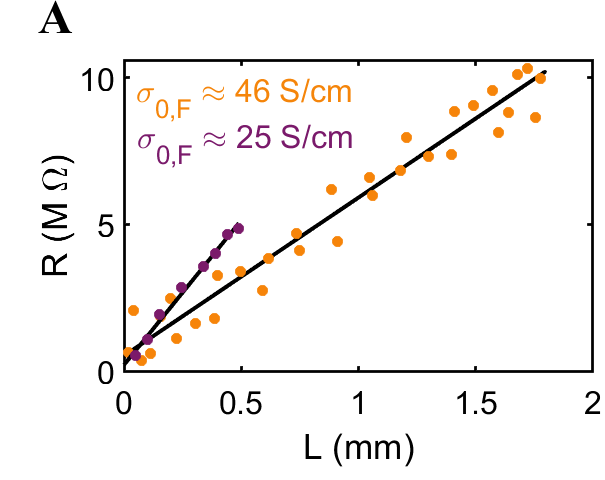}
    \end{subfigure}
    \begin{subfigure}[t]{0.45\textwidth}
        \raggedright
        \includegraphics[scale=0.37]{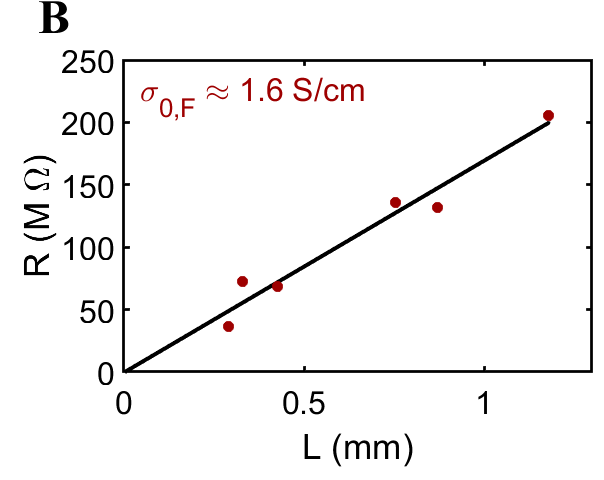}
\end{subfigure}
\caption{\textbf{Linear dependence of electrical resistance on probing length over mm distances.} (\textbf{A} and \textbf{B}) Four-probe measurements of the resistance, $R$, as a function of the electrode spacing, $L$, at 300 K for three separate fiber sheath filaments of cable bacteria. In these measurements, the two current-carrying electrodes and one voltage-sensing electrode were kept at a fixed position, while the second voltage-sensing electrode was varied along the filament, for increasing $L$. The resistance was calculated as $R={V}/{I}$, where $V$ is the voltage measured across the voltage-sensing electrodes and $I$ is the applied current. The resistance increases linearly with the probing length $L$, thus demonstrating that the fiber network possesses a uniform conductivity. The black lines represent linear fits through the data, which provide an estimate of the fiber conductivity (inset values) derived as $\sigma_{\rm 0,F}= 4 /(N_{\rm F} \pi {d_{\rm F}}^2 b)$, where $b$ is the fitted slope, $N_{\rm F}$ = 60 is the number of fibers in a filament (as determined by microscopy), and $d_{\rm F}=$ 26 nm is the cross-sectional diameter of the conductive core of a fiber \cite{meysman2019highly,boschker2021efficient}.}
\label{fig:EXT_FIG_LengthDep}
\end{figure}

\newpage
\subsection*{Figure S3}

\begin{figure}[h!]
    \centering

    \vspace{0.1cm}
    \begin{subfigure}[t]{0.44\textwidth}
        \raggedleft
        \includegraphics[scale=0.37]{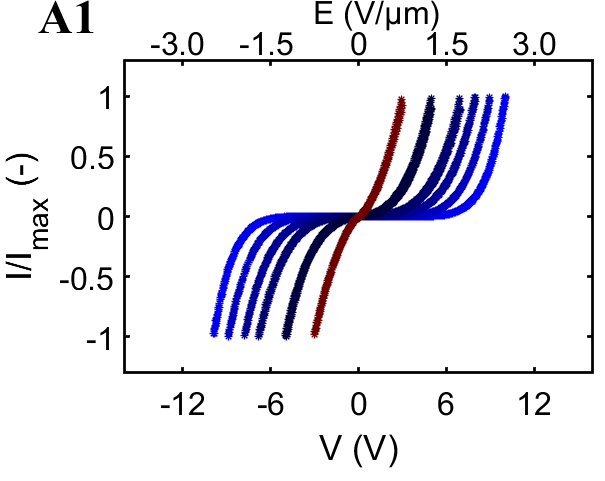}
    \end{subfigure}
    \begin{subfigure}[t]{0.44\textwidth}
        \includegraphics[scale=0.37]{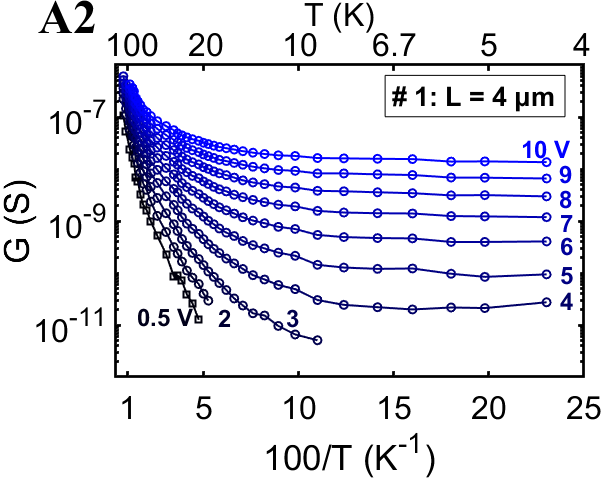}
    \end{subfigure}
    
    \vspace{0.1cm}
    \begin{subfigure}[t]{0.44\textwidth}
         \raggedleft
        \includegraphics[scale=0.37]{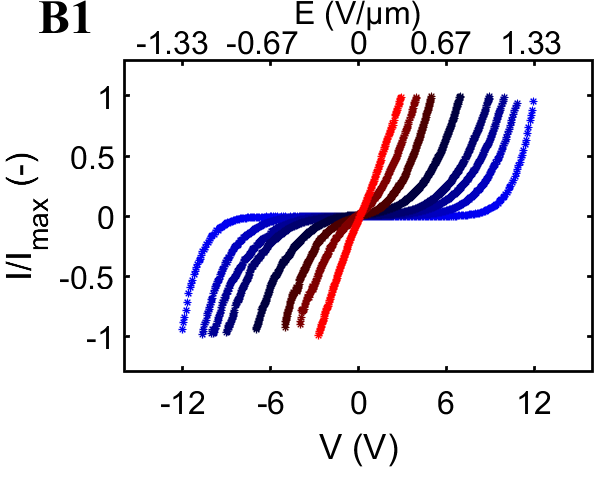}
    \end{subfigure}
    \begin{subfigure}[t]{0.44\textwidth}
        \includegraphics[scale=0.37]{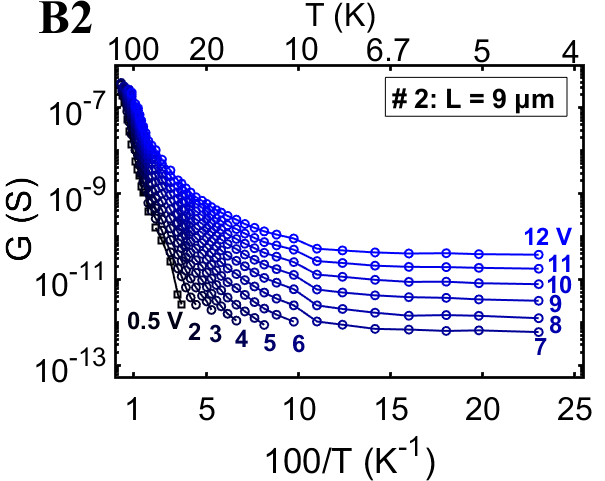}
    \end{subfigure}
    
    \vspace{0.1cm}
    \begin{subfigure}[t]{0.44\textwidth}
         \raggedleft
        \includegraphics[scale=0.37]{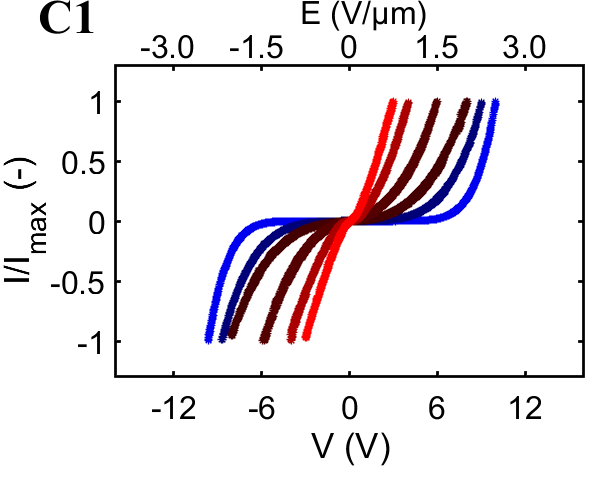}
    \end{subfigure}
    \begin{subfigure}[t]{0.44\textwidth}
        \includegraphics[scale=0.37]{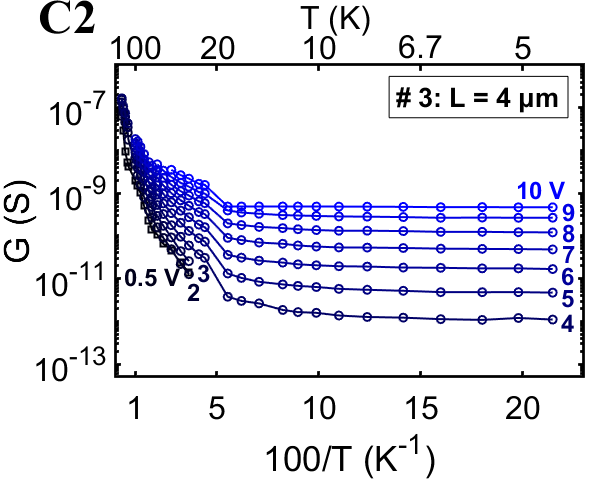}
    \end{subfigure}
            
    \vspace{0.1cm}
    \begin{subfigure}[t]{0.44\textwidth}
         \raggedleft
        \includegraphics[scale=0.37]{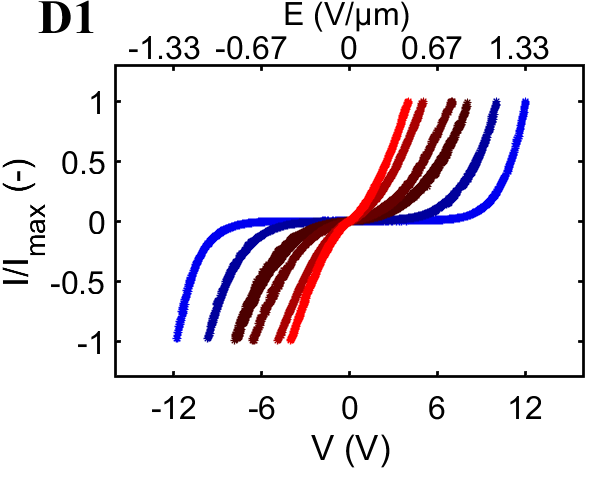}
    \end{subfigure}
    \begin{subfigure}[t]{0.44\textwidth}
        \includegraphics[scale=0.37]{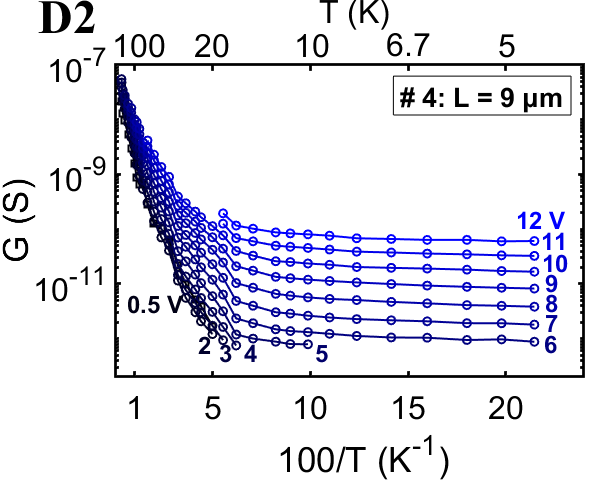}
    \end{subfigure}
\end{figure}

\newpage
\begin{figure}[h!]
    \centering
    \ContinuedFloat

    \vspace{0.1cm}
    \begin{subfigure}[t]{0.44\textwidth}
        \raggedleft        
        \includegraphics[scale=0.37]{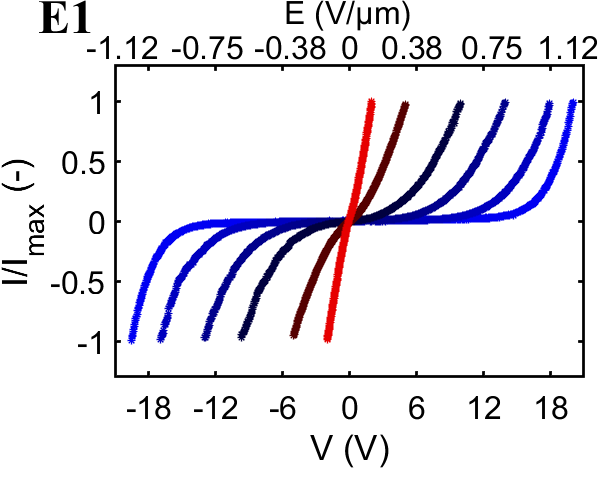}
    \end{subfigure}
    \begin{subfigure}[t]{0.44\textwidth}
      \includegraphics[scale=0.37]{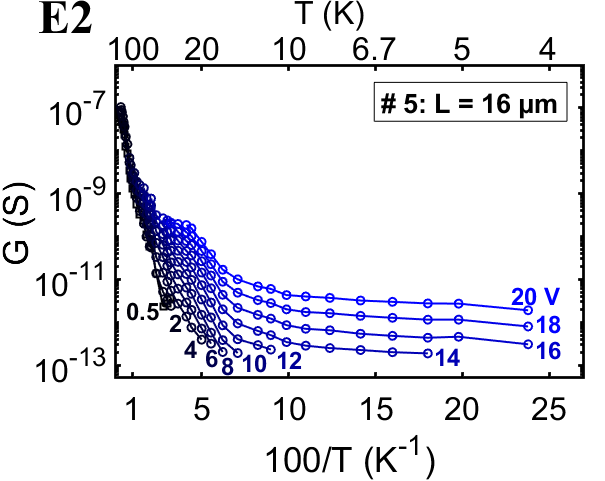}
    \end{subfigure}

    \vspace{0.1cm}
    \begin{subfigure}[t]{0.44\textwidth}
         \raggedleft
        \includegraphics[scale=0.37]{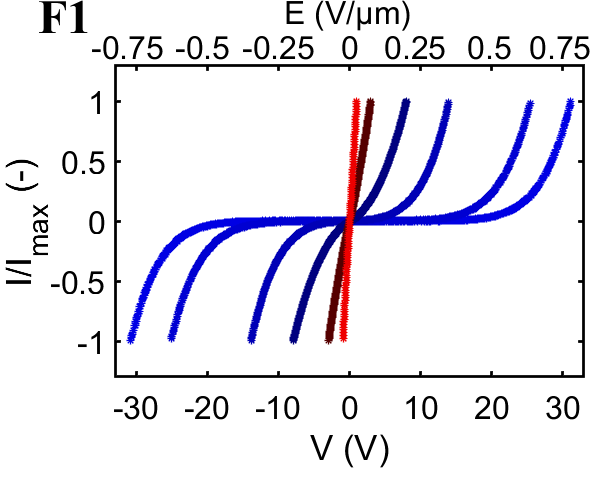}
    \end{subfigure}
    \begin{subfigure}[t]{0.44\textwidth}
        \includegraphics[scale=0.37]{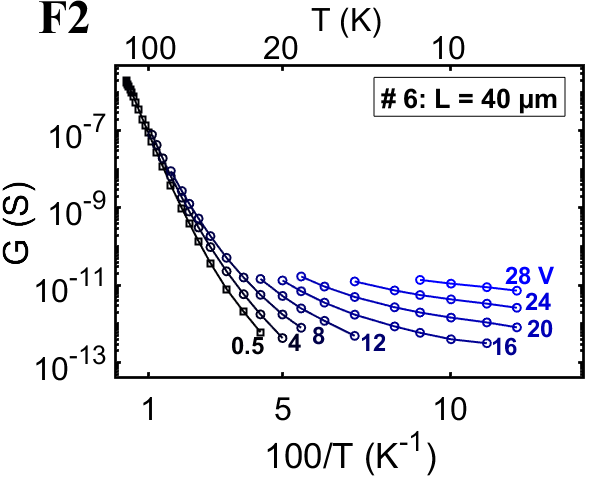}
    \end{subfigure}
    
    \vspace{0.1cm}
    \begin{subfigure}[t]{0.44\textwidth}
         \raggedleft
        \includegraphics[scale=0.37]{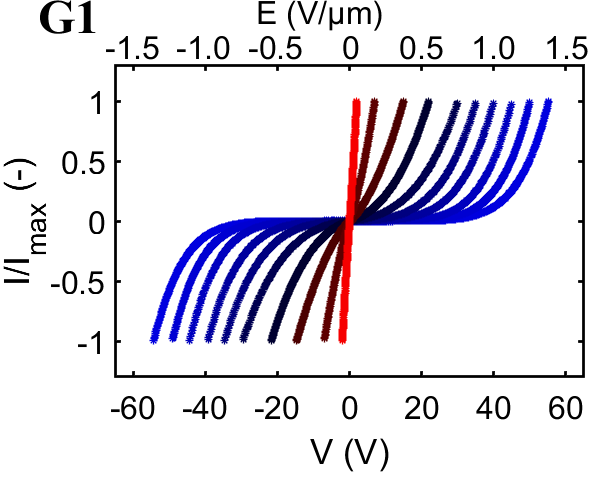}
    \end{subfigure}
    \begin{subfigure}[t]{0.44\textwidth}
        \includegraphics[scale=0.37]{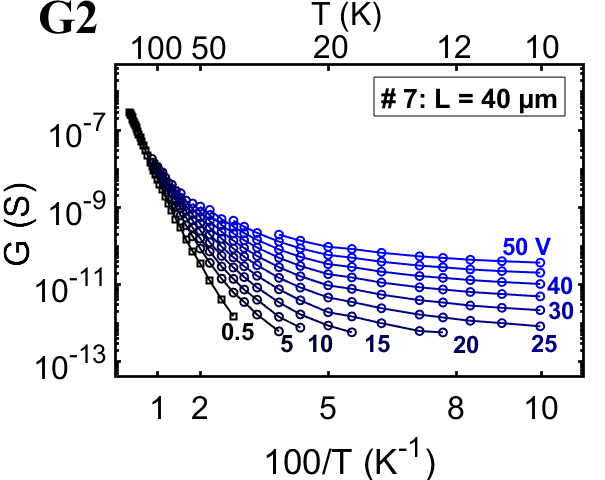}
    \end{subfigure}

    \vspace{0.1cm}
    \begin{subfigure}[t]{0.44\textwidth}
        \raggedleft
        \includegraphics[scale=0.37]{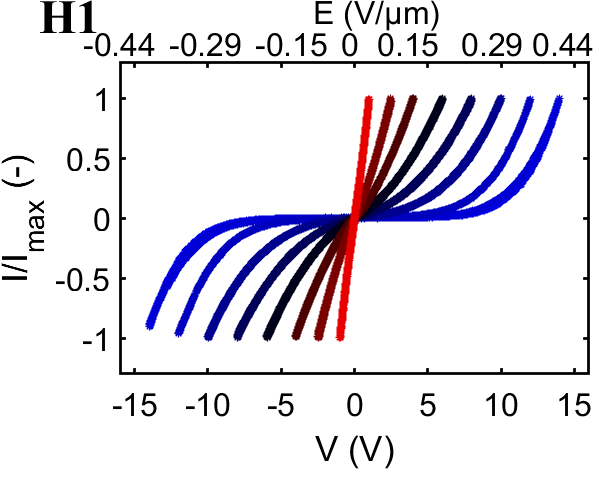}
    \end{subfigure}
    \begin{subfigure}[t]{0.44\textwidth}
        \includegraphics[scale=0.37]{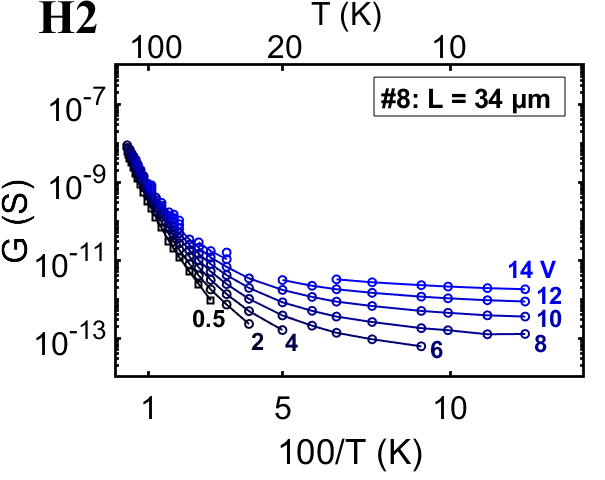}
    \end{subfigure}
\end{figure}
    
\newpage
\begin{figure}[h!]
    \centering
    \ContinuedFloat

    \vspace{0.1cm}
    \begin{subfigure}[t]{0.44\textwidth}
         \raggedleft
        \includegraphics[scale=0.37]{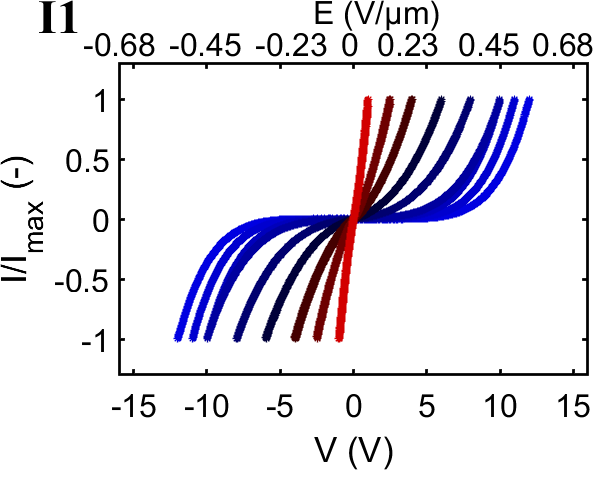}
    \end{subfigure}
    \begin{subfigure}[t]{0.44\textwidth}
        \includegraphics[scale=0.37]{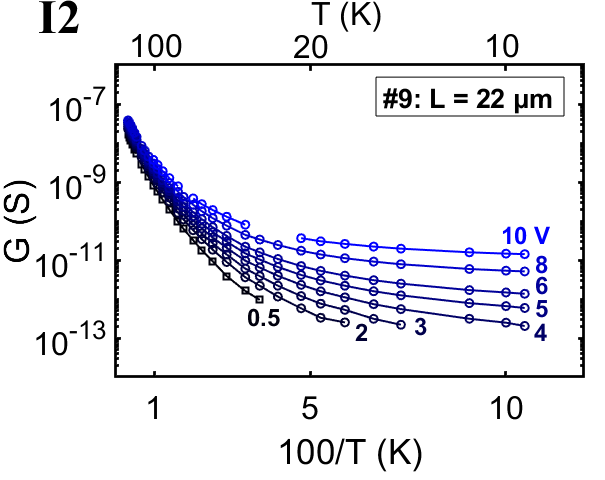}
    \end{subfigure}
    
    \vspace{0.1cm}
    \begin{subfigure}[t]{0.44\textwidth}
         \raggedleft
        \includegraphics[scale=0.37]{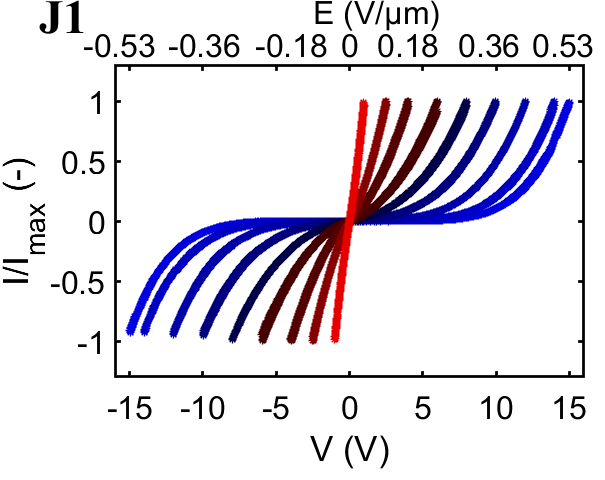}
    \end{subfigure}
    \begin{subfigure}[t]{0.44\textwidth}
        \includegraphics[scale=0.37]{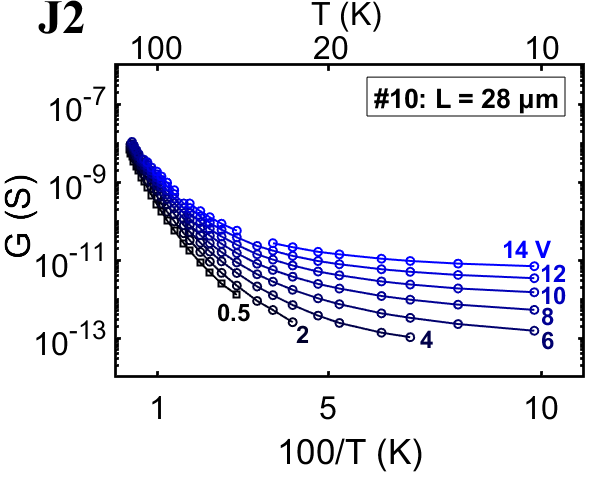}
    \end{subfigure}

    \vspace{0.1cm}
    \begin{subfigure}[t]{0.44\textwidth}
         \raggedleft
        \includegraphics[scale=0.37]{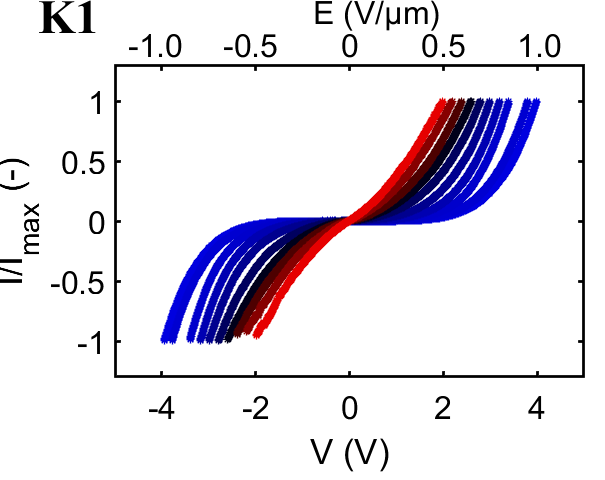}
    \end{subfigure}
    \begin{subfigure}[t]{0.44\textwidth}
        \includegraphics[scale=0.37]{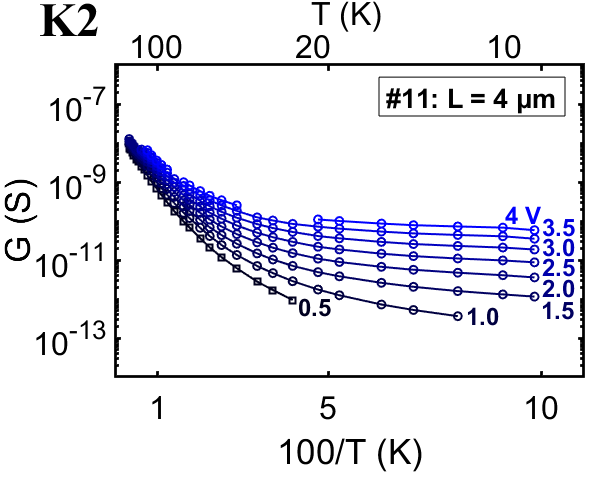}
    \end{subfigure}

    \vspace{0.1cm}
    \begin{subfigure}[t]{0.44\textwidth}
        \raggedleft        
        \includegraphics[scale=0.37]{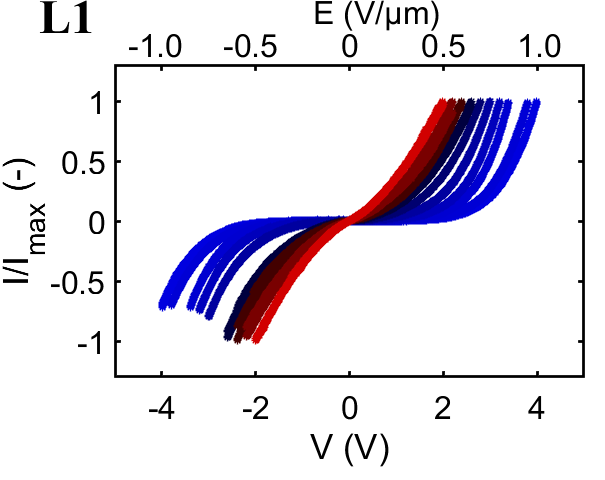}
    \end{subfigure}
    \begin{subfigure}[t]{0.44\textwidth}
      \includegraphics[scale=0.37]{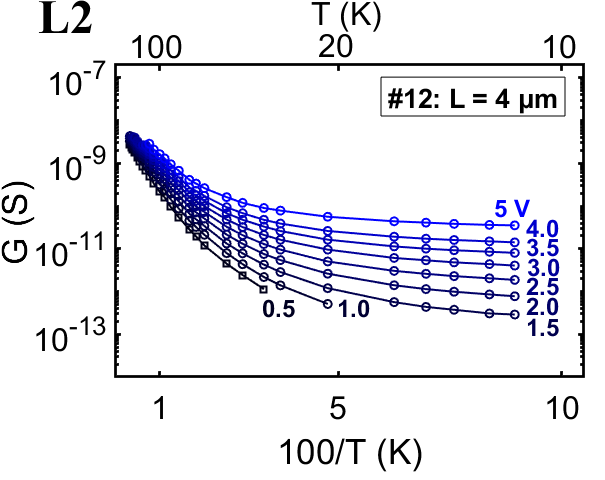}
    \end{subfigure}

\end{figure}

\newpage
\begin{figure}[h!]
    \centering    
    \ContinuedFloat

    \vspace{0.1cm}
    \begin{subfigure}[t]{0.44\textwidth}
         \raggedleft
        \includegraphics[scale=0.37]{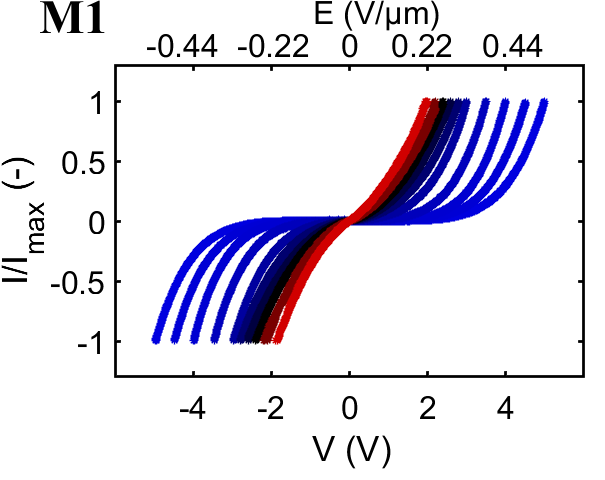}
    \end{subfigure}
    \begin{subfigure}[t]{0.44\textwidth}
        \includegraphics[scale=0.37]{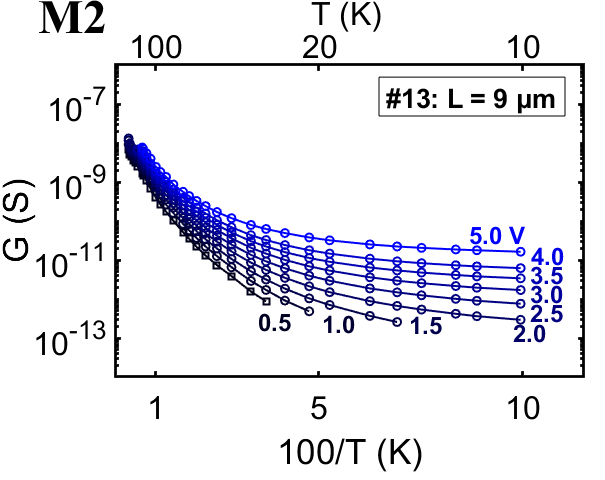}
    \end{subfigure}
    
    \vspace{0.1cm}
    \begin{subfigure}[t]{0.44\textwidth}
         \raggedleft
        \includegraphics[scale=0.37]{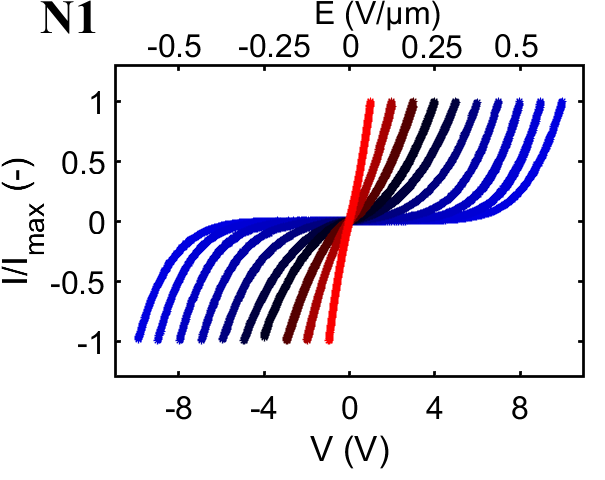}
    \end{subfigure}
    \begin{subfigure}[t]{0.44\textwidth}
        \includegraphics[scale=0.37]{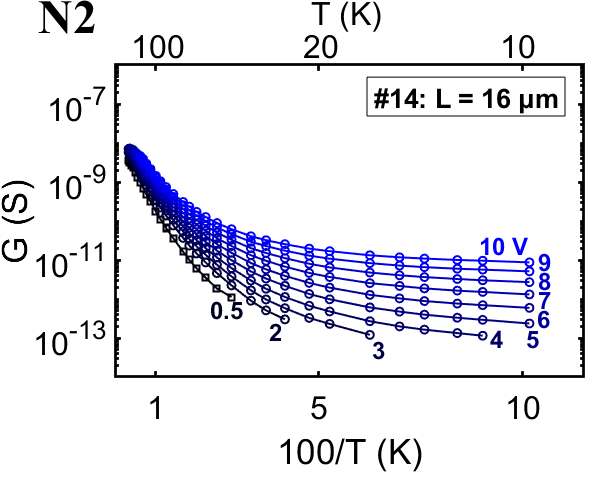}
    \end{subfigure}
    
    \vspace{0.1cm}
    \begin{subfigure}[t]{0.44\textwidth}
         \raggedleft
        \includegraphics[scale=0.37]{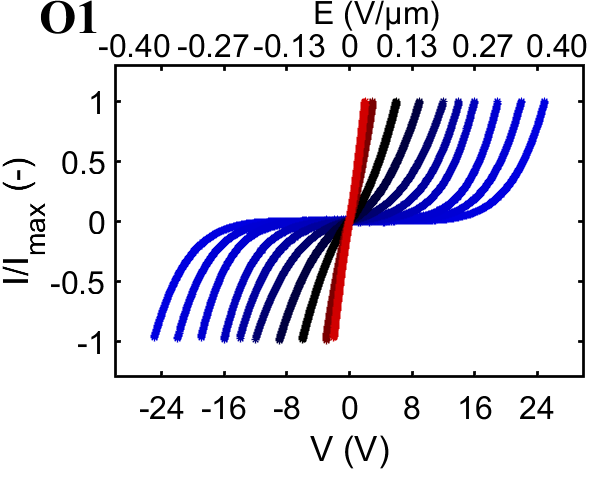}
    \end{subfigure}
    \begin{subfigure}[t]{0.44\textwidth}
        \includegraphics[scale=0.37]{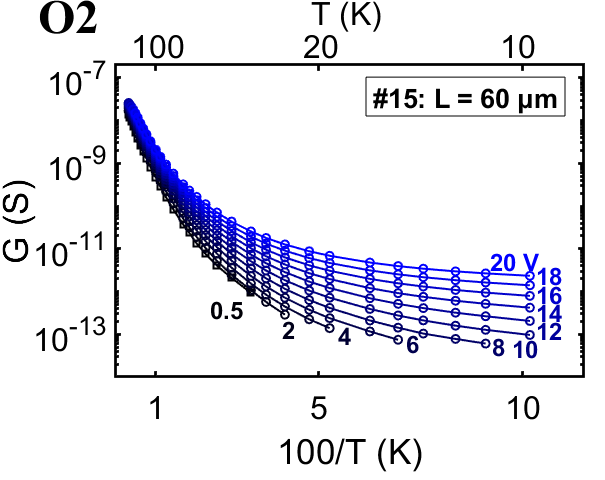}
    \end{subfigure}
\end{figure}   

\newpage
\begin{figure}[h!]
    \ContinuedFloat
    \vspace{0.1cm}
    \begin{subfigure}[t]{0.44\textwidth}
         \raggedleft
        \includegraphics[scale=0.37]{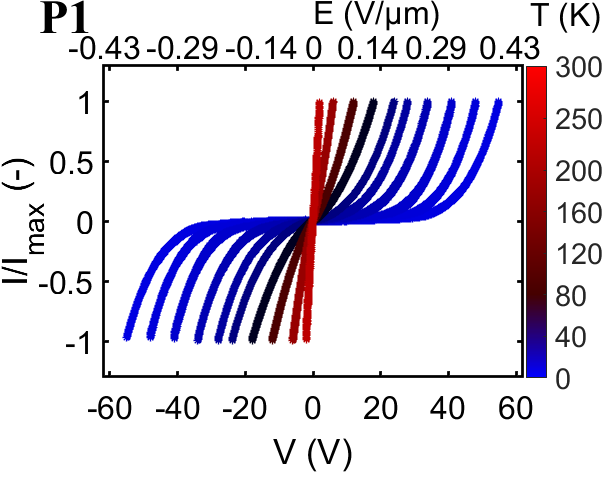}
    \end{subfigure}
    \begin{subfigure}[t]{0.44\textwidth}
        \includegraphics[scale=0.37]{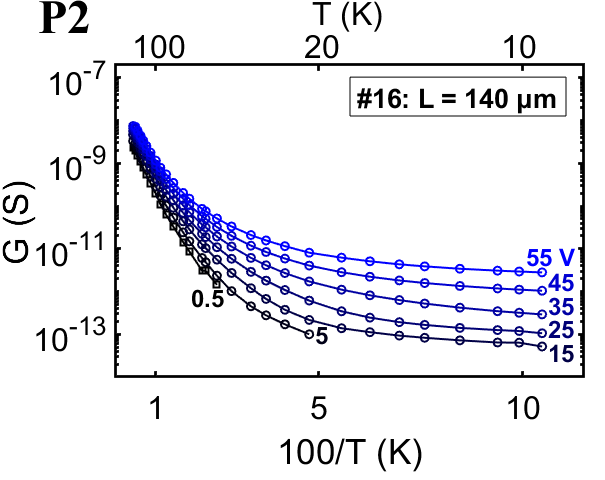}
    \end{subfigure}
    \caption{\textbf{Current/voltage measurements as a function of temperature.} The two panels in each row display data for a given fiber sheath segment of length $L$. Data are provided for 16 segments of different length. (\textbf{A1} to \textbf{P1}) Two-probe current-voltage ($I(V)$) characteristics recorded at different temperatures. To illustrate the change in shape of the $I(V)$ curve, the current is normalized to the maximum current obtained for the given trace. The top axis shows the applied electric field, $E=V/L$. The colour of the lines indicates the temperature (linearly changing from red to black in the range $300 - 80$ ~K and from black to blue in the range $80 - 4$~K). See colour bar next to panel (P1). (\textbf{A2} to \textbf{P2}) The same dataset is replotted in terms of the conductance, $G=I/V$, as a function the inverse temperature, $100/T$, for different bias voltages, $V$. The colour scale linearly ranges from black (zero-bias voltage) to blue (maximum bias voltage). Note that this voltage range differs per segment. The segment number and segment length are indicated in the right panels (see also table~S\ref{table:basic}).}
    \label{fig:IV_GT_All}
\end{figure}

\newpage
\subsection*{Figure S4}

\begin{figure}[h!]
    \centering
    \begin{subfigure}[t]{0.44\textwidth}
        \raggedleft
        \includegraphics[scale=0.37]{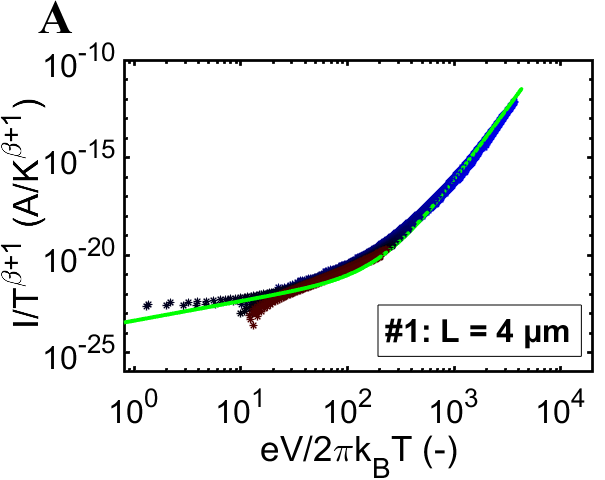}
    \end{subfigure}
    \begin{subfigure}[t]{0.44\textwidth}
        \includegraphics[scale=0.37]{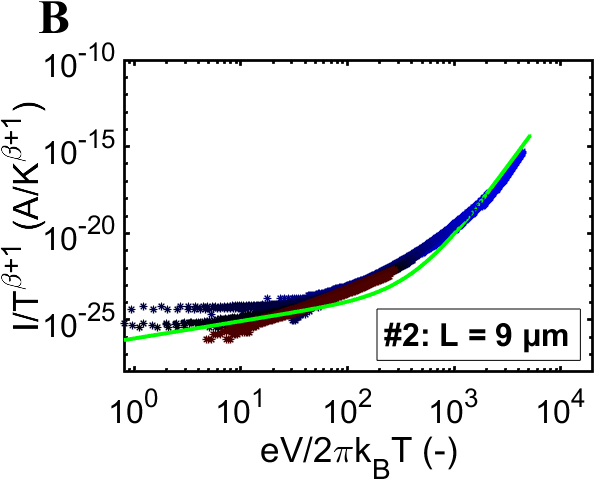}
    \end{subfigure}
    
    \vspace{0.1cm}
    \begin{subfigure}[t]{0.44\textwidth}
        \raggedleft
        \includegraphics[scale=0.37]{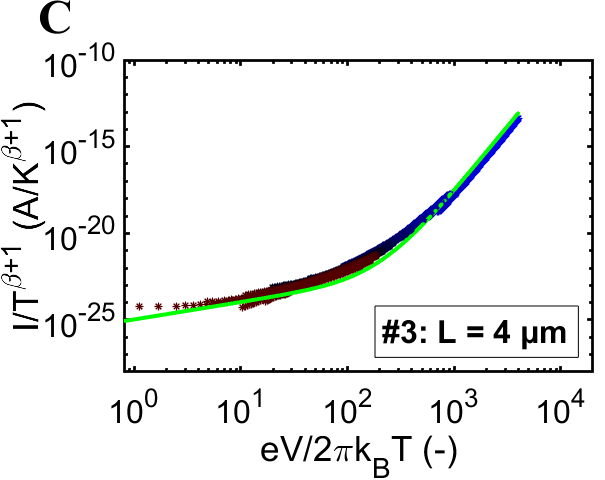}
    \end{subfigure}
    \begin{subfigure}[t]{0.44\textwidth}
        \includegraphics[scale=0.37]{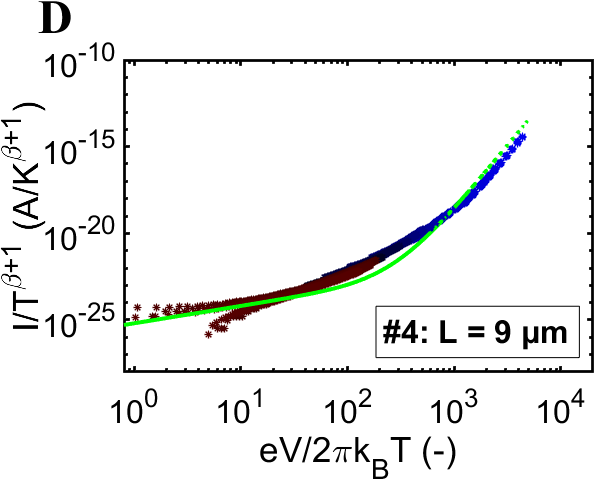}
    \end{subfigure}
    
    \vspace{0.1cm}
    \begin{subfigure}[t]{0.44\textwidth}
        \raggedleft
        \includegraphics[scale=0.37]{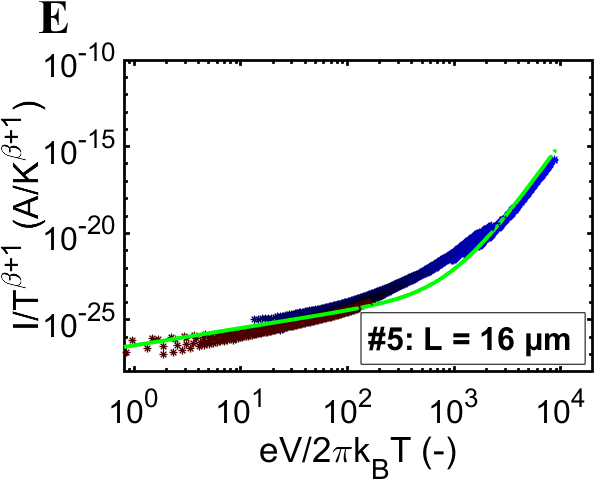}
    \end{subfigure}
    \begin{subfigure}[t]{0.44\textwidth}
        \includegraphics[scale=0.37]{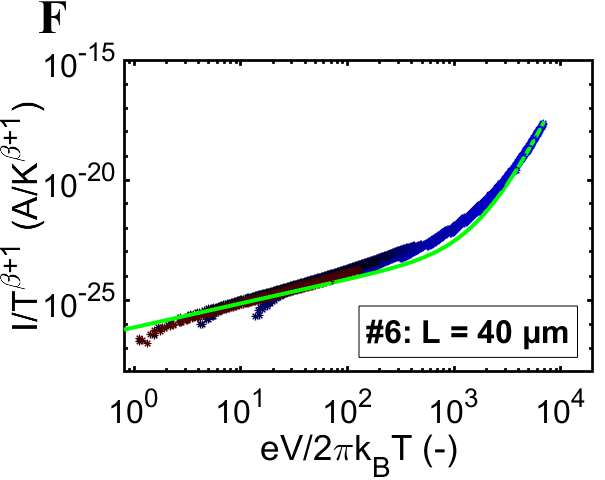}
    \end{subfigure}
    
    \vspace{0.1cm}
    \begin{subfigure}[t]{0.44\textwidth}
        \raggedleft
        \includegraphics[scale=0.37]{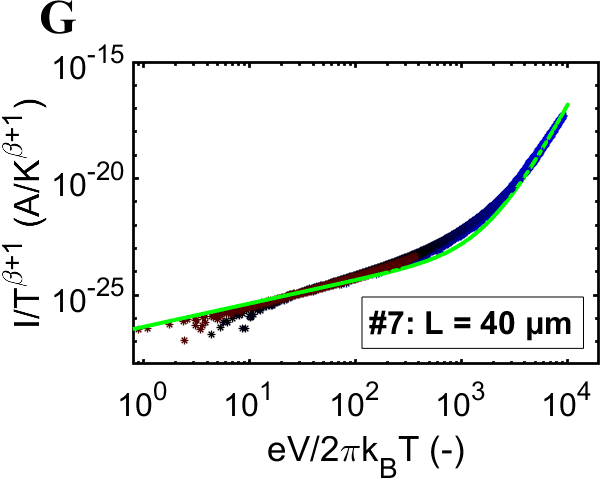}
    \end{subfigure}
    \begin{subfigure}[t]{0.44\textwidth}
        \includegraphics[scale=0.37]{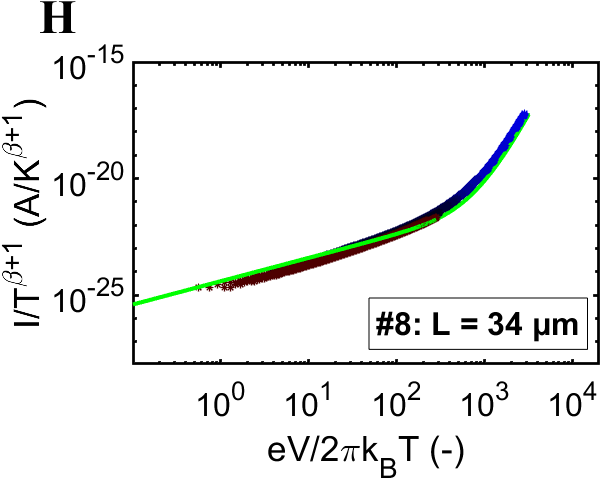}
    \end{subfigure}
\end{figure}   

\newpage
\begin{figure}[h!]
    \ContinuedFloat
    \centering
    \begin{subfigure}[t]{0.44\textwidth}
        \raggedleft
        \includegraphics[scale=0.37]{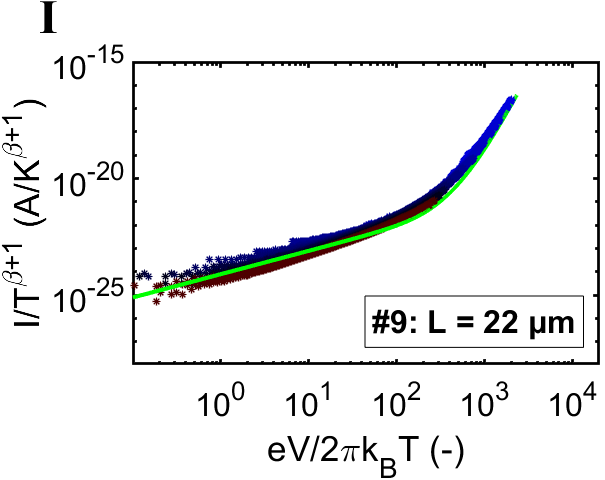}
    \end{subfigure}
    \begin{subfigure}[t]{0.44\textwidth}
        \includegraphics[scale=0.37]{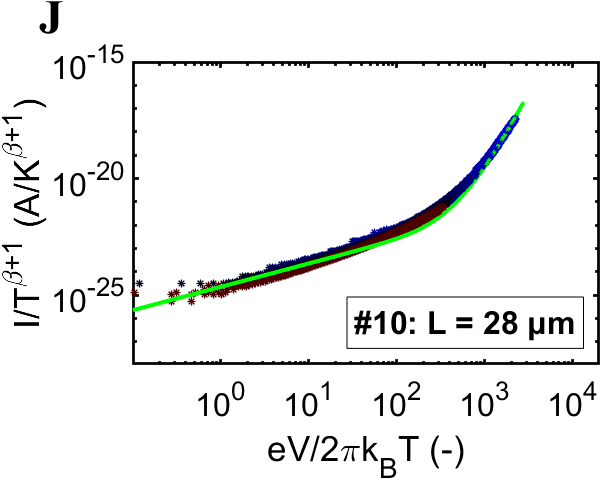}
    \end{subfigure}
    
    \vspace{0.1cm}
    \begin{subfigure}[t]{0.44\textwidth}
        \raggedleft
        \includegraphics[scale=0.37]{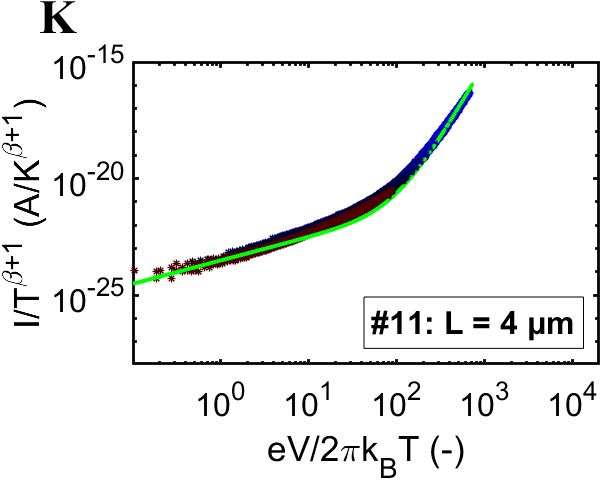}
    \end{subfigure}
    \begin{subfigure}[t]{0.44\textwidth}
        \includegraphics[scale=0.37]{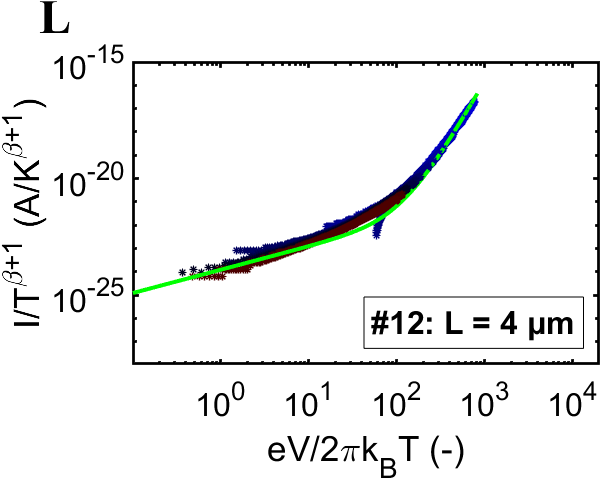}
    \end{subfigure}
    
    \vspace{0.1cm}
    \begin{subfigure}[t]{0.44\textwidth}
        \raggedleft
        \includegraphics[scale=0.37]{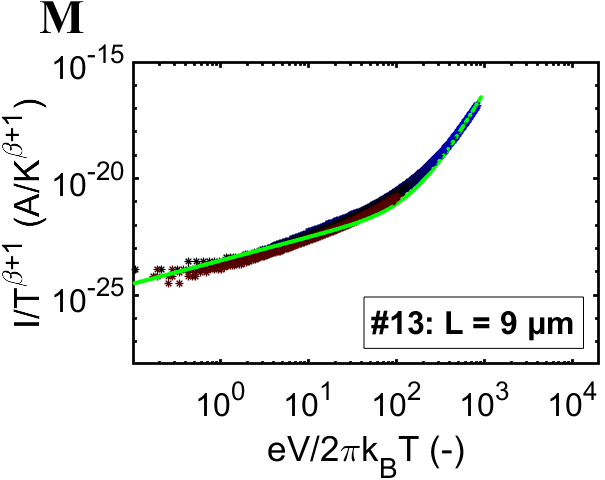}
    \end{subfigure}
    \begin{subfigure}[t]{0.44\textwidth}
        \includegraphics[scale=0.37]{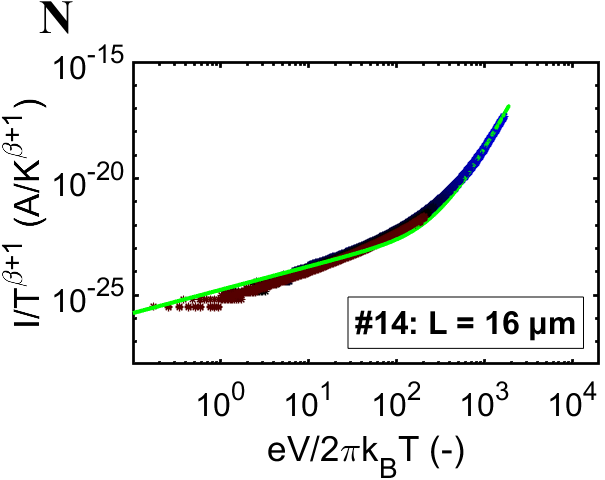}
    \end{subfigure}

\end{figure}   

\newpage
\begin{figure}[h!]
    \ContinuedFloat
    \centering
    \vspace{0.1cm}
    \begin{subfigure}[t]{0.44\textwidth}
        \raggedleft
        \includegraphics[scale=0.37]{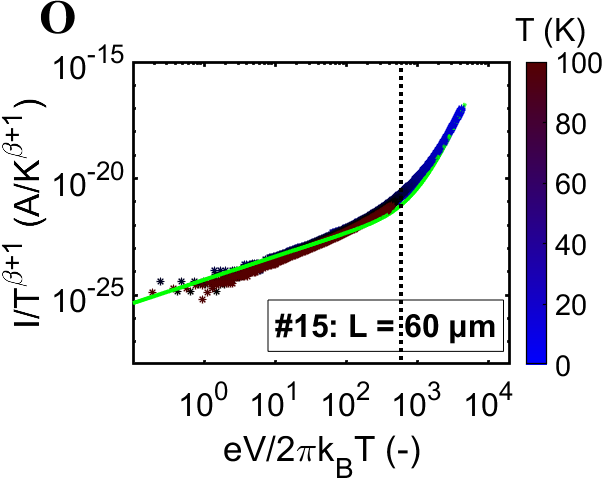}
    \end{subfigure}
    \begin{subfigure}[t]{0.44\textwidth}
        \includegraphics[scale=0.37]{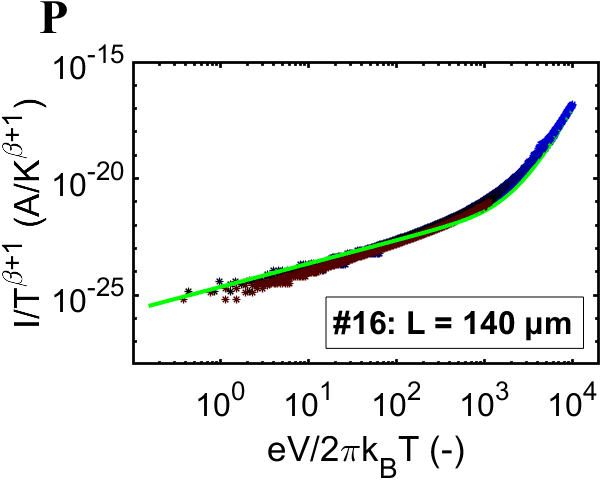}
    \end{subfigure}
    \caption{\textbf{Universal scaling behaviour of the conductance at low temperatures.} (\textbf{A} to \textbf{P}) Scaling curves are displayed for the 16 segments shown in  Fig.~\ref{fig:IV_GT_All}. The plots display the scaled current $I/T^{\beta+1}$ on the $y$-axis versus the scaled bias $eV/2\pi k_{\rm B} T$ on the $x$-axis. The $I(V)$ data in the temperature range below 100 K are used. The temperature at which individual data points were collected is indicated by the colour scale next to panel (\textbf{O}). The green solid line provides a fit of the universal scaling relation (Eq.~\ref{eq:USC}) to the data. The power law exponent $\beta$ is visually calibrated to obtain the best fit (precision $\pm \ 0.5$). The transition from the low-bias, high-temperature regime to the high-bias, low-temperature regime occurs at a bias $eV \approx 2\pi N_{\rm S} k_{\rm B} T$. This transition point hence determines the parameter $N_{\rm S}$, which signifies the apparent number of electron transfer steps in the hopping chain. In panel (\textbf{O}) this transition point is indicated by the dotted line. The resulting values for the power law exponent, $\beta$ and the apparent number of hopping sites, $N_{\rm S}$, are listed in table~S\ref{table:NT}. The parameter $N_{\rm S}$ scales with the segment length (see fig.~S\ref{fig:NL}).}
    \label{fig:USC_All}
\end{figure}

\clearpage
\newpage
\subsection*{Figure S5}

\begin{figure}[h!]
    \begin{subfigure}{0.32\textwidth}
      \includegraphics[scale=0.36]{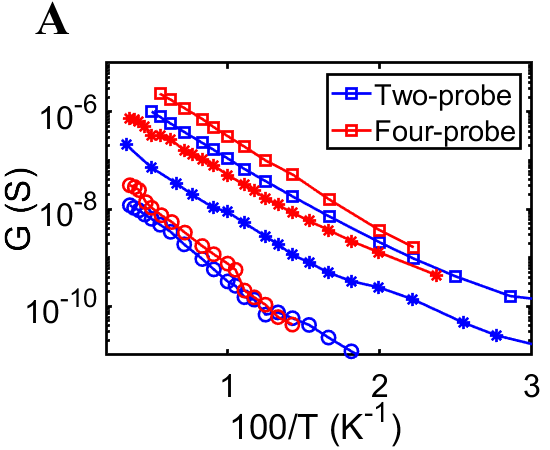}
    \label{fig:SI_1A}
    \end{subfigure}
    \begin{subfigure}{0.32\textwidth}
        \includegraphics[scale=0.36]{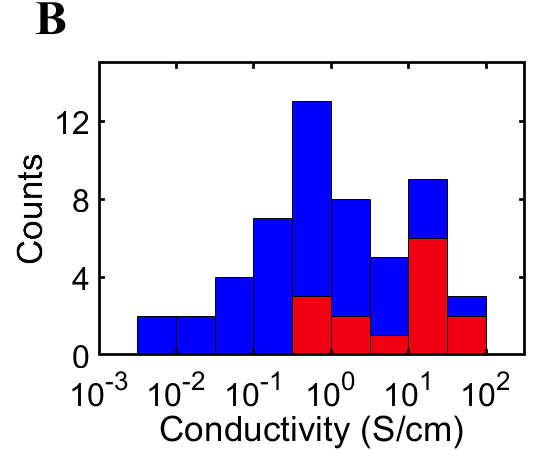}
    \label{fig:SI_1B}
    \end{subfigure}
    \begin{subfigure}{0.32\textwidth}
      \includegraphics[scale=0.36]{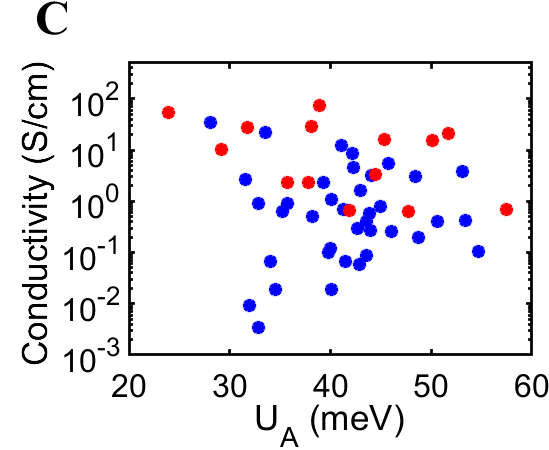}
     \label{fig:SI_1C}
    \end{subfigure}
    \caption{\textbf{High-temperature conductivity analysis.} (\textbf{A}) Temperature dependence of conductance, $G$, as measured by the two-probe and four-probe approach for three different segments. As expected, the four-probe conductance is systematically higher as it does not include the contact resistance. The two-probe and four-probe data show a very similar temperature dependence, demonstrating that both methods capture the same intrinsic temperature response of the conductance. The activation energy, $U_{\rm A}$, is determined as the slope from an Arrhenius plot, i.e., a linear fit of  $\text{ln}(G)$ vs. $1/(k_{\rm B} T)$ for $T > 100$ K (see also fig.~S\ref{fig:Arrhenius}). We find similar activation energies for the two-probe and four-probe data: first segment (squares):  $38.1 \pm 0.2$ meV vs. $38.2 \pm 2$ meV; second segment (stars): $41.2\pm 6.7$ meV vs. $35.7\pm 2.3$ meV; third segment (circles):  $48.5 \pm 2.3$ meV vs. $50.6 \pm 3.6$ meV. (\textbf{B}) Histogram of the room-temperature conductivity for all fiber sheath segments investigated ($n=53$). Conductivity was calculated via Eq.~\ref{eq:conductivity} as detailed in the Supplementary Text. Both four-probe (red bars) and two-probe (blue bars) estimates are shown. Note, that four-probe values are typically higher than two-probe values, due to the absence of the contact resistance. (\textbf{C}) Activation energy $U_{\rm A}$ of all segments investigated ($n$ = 53) plotted against the room-temperature fiber conductivity, $\sigma_{\rm 0,F}$. Two-probe data (blue markers) and four-probe data (red markers) are provided. The activation energy, $U_{\rm A}$, and the room temperature conductivity, $\sigma_{0,F}$, are uncorrelated (Pearson correlation coefficient =  0.29). While the variance in $\sigma_{0,F}$ is rather large, the variation in $U_{\rm A}$ is small.}
    \label{fig:HighT_G_Analysis}
\end{figure}

\clearpage
\newpage
\subsection*{Figure S6}

\begin{figure}[h]
    \centering
    \begin{subfigure}[t]{0.45\textwidth}
        \raggedleft
        \includegraphics[scale=0.37]{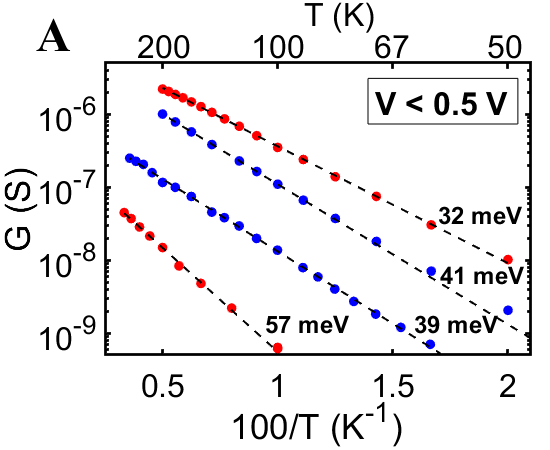}
    \end{subfigure}
    \begin{subfigure}[t]{0.45\textwidth}
        \raggedright
        \includegraphics[scale=0.37]{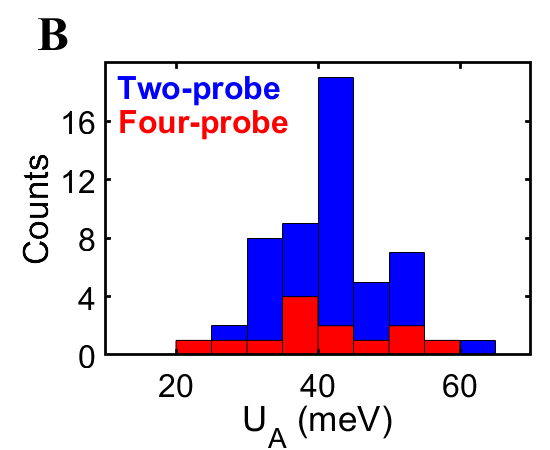}
\end{subfigure}
\caption{\textbf{Arrhenius temperature dependence of the conductance.} (\textbf{A}) Low-bias conductance as a function of temperature plotted above 50~K for two- and four-probe measurements, shown in blue and red respectively. Measurements taken on four different segments. Arrhenius fits (performed for data where $T > 100$ K) and corresponding activation energies are indicated. (\textbf{B}) Histogram with the activation energies measured in $n = 53$ segments (blue: two-probe measurements; red: four-probe measurements). The mean activation energy is $42 \pm 8$ meV ($n=$ 53) and the mean goodness of fit across all samples as expressed by the coefficient of determination is $\langle R^2 \rangle \approx 0.982$.}
\label{fig:Arrhenius}
\end{figure}

\clearpage
\newpage
\subsection*{Figure S7}

\begin{figure}[h!]
\centering
\includegraphics[scale=0.3]{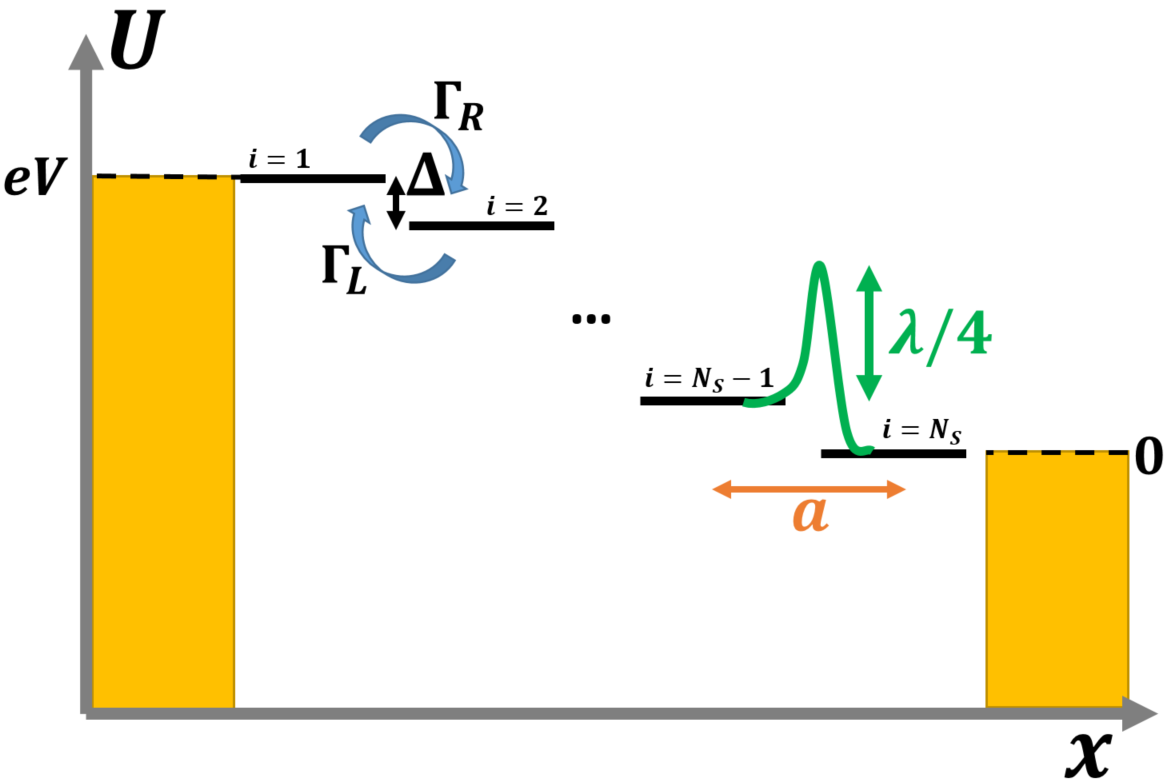}
\caption{\textbf{Schematic of a one-dimensional hopping chain.} The vertical axis denotes the energy, $U$, and the horizontal axis the position, $x$, along the chain. The yellow rectangles represent the electrodes that have an applied voltage bias $V$ ($U=0$ at the right electrode and $U = eV$ at the left electrode). The horizontal black lines indicate the charge carrier sites, numbered from $i=1$ to $i=N_{\rm S}$,  where $N_{\rm S}$ is the number of sites. The parameter $a$ is the center-to-center distance between adjacent sites. There is an equal energy drop between neighboring sites, $\Delta = eV/N_{\rm S}$. There are forward hopping rates, $\Gamma_{\rm R}$, and backwards hopping rates, $\Gamma_{\rm L}$, between any pairs of sites, but only hopping rates between nearest neighbour sites are taken into account. The reorganization energy, $\lambda$, determines the activation barrier for hopping (see Supplementary Text).}
\label{fig:HoppingSchematic}
\end{figure}

\newpage
\subsection*{Figure S8}

\begin{figure}[h!]
    \centering
    \begin{subfigure}[t]{0.45\textwidth}
        \raggedleft
        \includegraphics[scale=0.38]{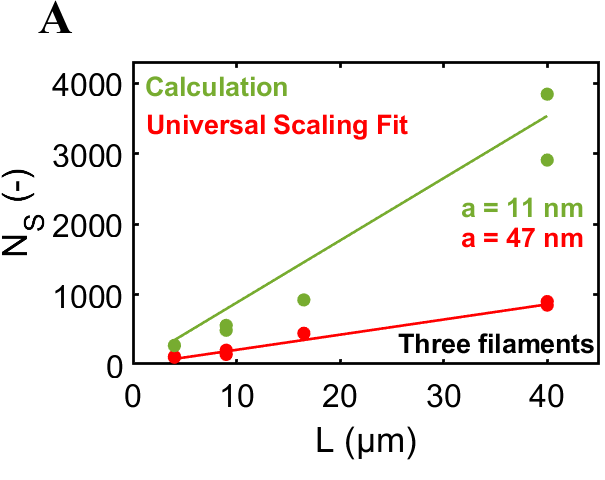}
    \end{subfigure}
    \begin{subfigure}[t]{0.45\textwidth}
        \raggedright
        \includegraphics[scale=0.38]{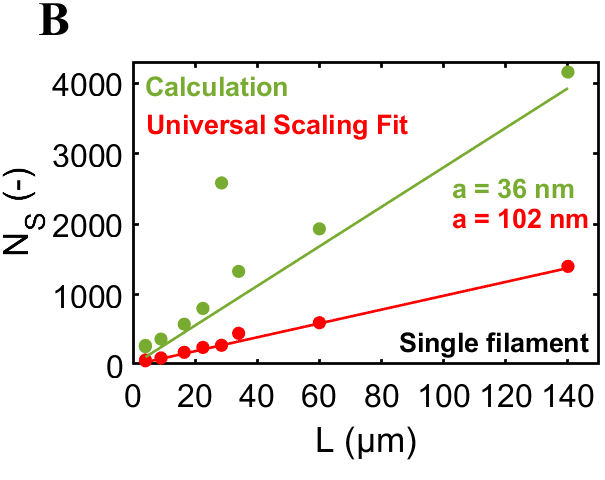}
\end{subfigure}
\caption{\textbf{Linear scaling of the number of hopping sites with the segment length.} The number of hopping steps, $N_{\rm S}$, is determined in two ways. In a first approach (red markers), $N_{\rm S}$ is derived from the transition point in the universal scaling curves (see fig.~S\ref{fig:USC_All}). In a second approach (green markers), $N_{\rm S}$ is calculated via Eq.~\ref{eq:NS_BASIC} in the Supplementary Text. In both cases, $N_{\rm S}$ linearly scales with the segment length $L$ (solid lines provide linear fits, excluding offset). The center-to-center distance between hopping sites, $a$, is given by the slope $a=L/N_{\rm S}$. Both methods imply $a$ values that are in excess of 10 nm. (\textbf{A}) Data from three independent fiber sheaths. (\textbf{B}) Data obtained from different segments on a single fiber sheath.}
\label{fig:NL}
\end{figure}

\newpage
\subsection*{Figure S9}

\begin{figure}[h!]
\centering
\includegraphics[scale=0.6]{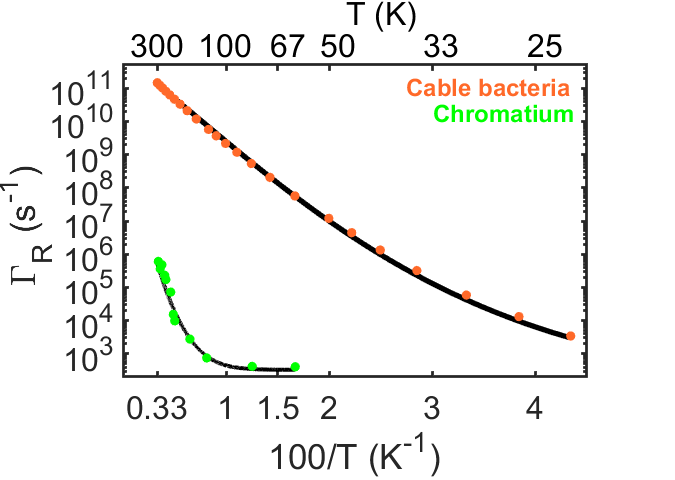}
\caption{\textbf{Comparison of electron transfer rates in the purple bacterium \emph{Allochromatium vinosum} versus the cable bacterium \emph{Electrothrix gigas}.} Experimental data are respectively obtained from Ref. 18 for \emph{A. vinosum} (green dots) and segment 6 (L = 40 $\mu$m) from the dataset provided here (green dots). The  hopping model with a single effective vibronic mode (Jortner model; Suppl. Text) is fitted to both datasets (solid lines). Fitting parameters for \emph{A. vinosum}: $\hbar\langle\omega\rangle = 60$ meV, $\lambda = 1.2$ eV and $V_{\rm DA} = 1$ meV. For \emph{E. gigas}, the measured current was first converted into an electron transfer rate via Eq.~\ref{eq:Ihigh} with parameters: voltage bias $V=0.5$ V, $N_{\rm S} = 850$, $N_{\rm F}=60$ and $N_{\rm C}=125$. The parameters used for the Jortner model fit are: $\hbar\langle\omega\rangle = 9$ meV, $\lambda = 0.24$ eV and $V_{\rm DA} = 6$ meV.}
\label{fig:ChromatiumCBJortner}
\end{figure}

\newpage
\section*{Tables S1 and S2}
\subsection*{Table S1}

\begin{table}[h]
\centering
\includegraphics[scale=0.370]{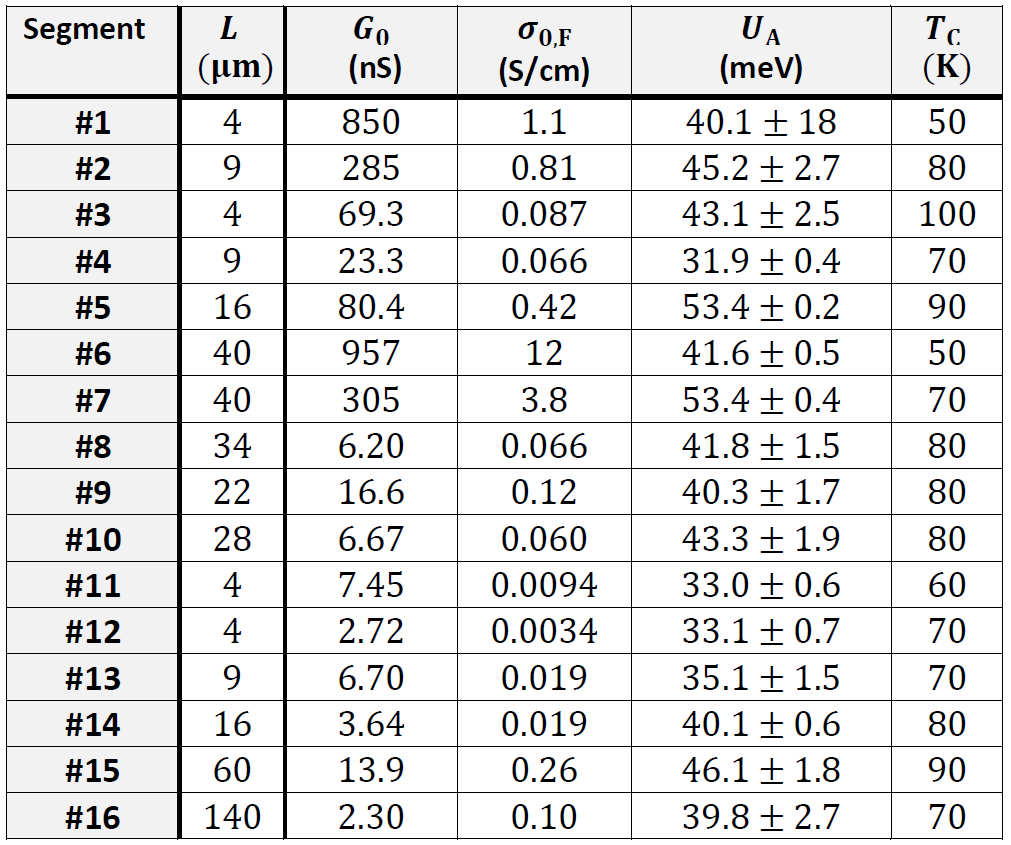}
\caption{\textbf{Parameters estimated from high-temperature conductance data}. The first column displays the sample identification number (16 segments have been analysed). $L$ denotes the segment length probed between two gold electrode pads. $G_0$ is the conductance measured at room temperature, $T= 300$~K. $\sigma_{\rm 0,F}$ is the fiber conductivity at $T= 300$ K calculated via Eq.~\ref{eq:conductivity}. The activation energy, $U_{\rm A}$, is determined by fitting the Arrhenius relation (Eq.~\ref{eq:Arrhenius}) to the high-temperature conductance data ($T >$ 100 K; see fig.~S\ref{fig:HighT_G_Analysis}). The cross-over temperature, $T_{\rm C}$, is the first temperature where the conductance as predicted by the Arrhenius fit is 20\% lower than the actually measured conductance.}
\label{table:basic}
\end{table}

\newpage
\subsection*{Table S2}
\begin{table}[h!]
\centering
\includegraphics[scale=0.5]{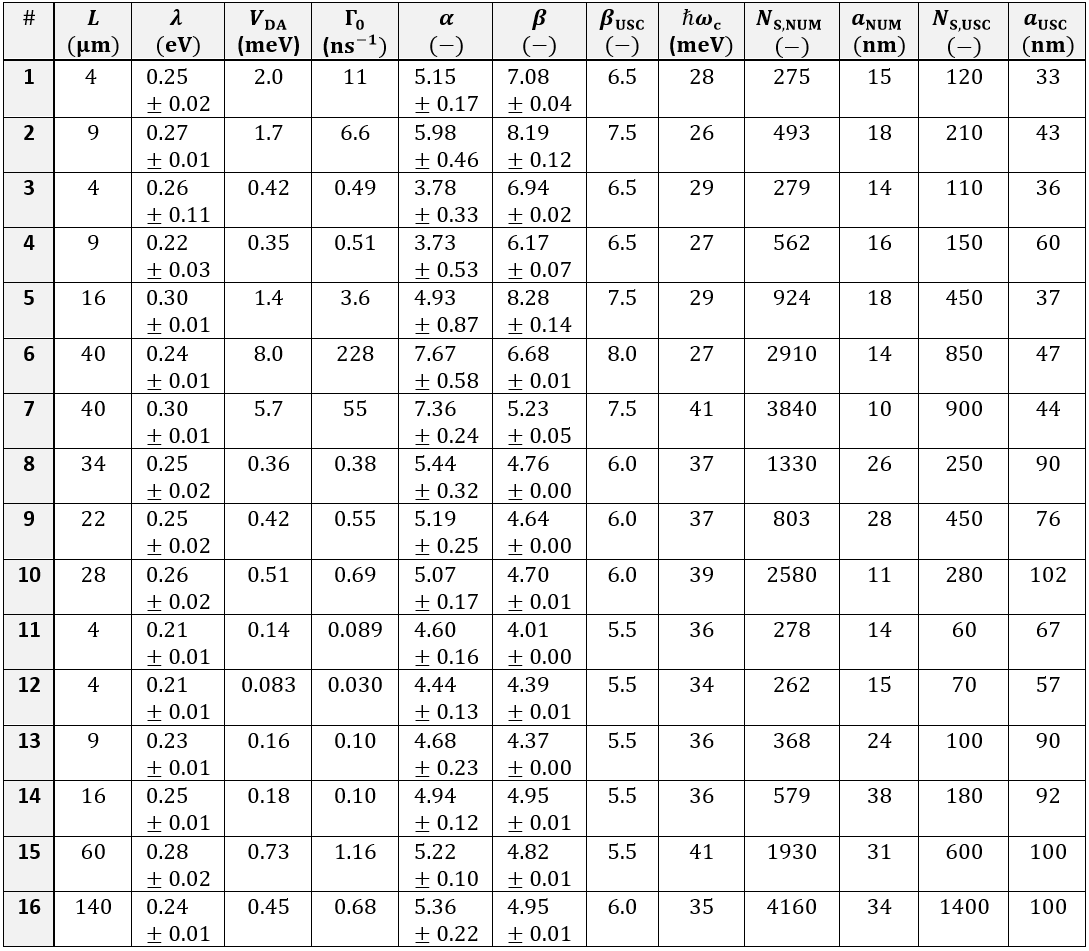}
\caption{\small \textbf{Parameters estimated from fitting the proposed hopping model.} The first column displays the sample identification number (same 16 segments as in table~S\ref{table:basic}). $L$ is the segment length between two electrode pads. The reorganization energy, $\lambda$, is obtained by fitting the Marcus expression (Eq.~\ref{eq:MarcusG}) to the high-temperature conductance data ($T$ > 100 K; see Fig.~\ref{fig:4}A). The electronic coupling, $V_{\rm DA}$, is obtained via Eq. \ref{eq:Marcus_VDA}. The transition rate, $\Gamma_0$, at 300 K is calculated from the Marcus rate expression (Eq.~\ref{eq:MarcusRate}) using the values for $\lambda$ and $V_{\rm DA}$ listed. The power law exponent, $\alpha$, was obtained by fitting the temperature-based power law $G \propto T^{\alpha}$ to the conductance data in the region 20 - 100 K (Fig.~\ref{fig:4}C). The power law exponent, $\beta$, was obtained by fitting the electric field-based power law (Eq.~\ref{eq:LowTRate} and Eq.~\ref{eq:Ilow}) to the conductance data for $T<10$~K (Fig.~\ref{fig:4}D). An alternative value for this exponent, $\beta_{\rm USC}$, is obtained by fitting the universal scaling curve (fig.~S\ref{fig:USC_All}). The characteristic vibration energy, $\hbar\omega_{\rm c}$, was determined via Eq.~\ref{eq:hbar_omega_c}. The number of sites $N_{\rm S,NUM}$ was obtained by Eq.~\ref{eq:NS_BASIC} and the corresponding center-to-center distance is $a_{\rm NUM}=L/N_{\rm S,NUM}$. Alternatively, the number of sites $N_{\rm S,USC}$ was also independently obtained from the universal scaling curves (fig.~S\ref{fig:USC_All}) and the corresponding center-to-center distance between sites, $a_{\rm USC}=L/N_{\rm S,USC}$ is also given.}
\label{table:NT}
\end{table}

\newpage
\section*{Supplementary Text}

\subsection*{Calculation of the fiber conductivity}
\nolinenumbers

The fiber conductivity, $\sigma_{0,F}$, at room temperature ($T = $ 300 K) was calculated from the experimental I/V data as:
\begin{linenomath}
\begin{equation}
\sigma_{\rm 0,F} = G_0 \frac {L} {N_{\rm F} A_F} = G_0 \frac {4 L} {N_{\rm F} \pi d_{\rm F}^2}. 
\label{eq:conductivity}
\end{equation}
\end{linenomath}
In this, $G_0$ represents the low-bias conductance of a single fiber sheath as determined at $T = 300$ K from the slope of the I/V curve, $L$ is the length of the fiber sheath segment investigated as determined by microscopy, $A_F$ is the cross-sectional area of one fiber, and $N_{\rm F}$ is the total number of fibers embedded in the fiber sheath. Experimentally determined values for $G_0$  and $L$ are given for different segments in table~S\ref{table:basic}. The number of parallel fibers $N_{\rm F}$ is determined by scanning electron microscopy or atomic force microscopy on individual filaments\cite{cornelissen2018cell} and is here $N_{\rm F}$ = 60 (see fig.~S\ref{fig:EXT_SEM}). The cross-sectional area is taken the same value for all segments: $d_{\rm F}=$ 26 nm is the diameter of the conductive core of an individual fiber, which is invariant across different filaments\cite{boschker2021efficient}.

\nolinenumbers
\subsection*{Hopping model of electron transport in cable bacteria}

\subsubsection*{Geometry of the conductive network}
\nolinenumbers

Cable bacteria possess a set of $N_{\rm F}$ fiber structures in their cell envelope, which are arranged in parallel and run continuously along the whole length of cm-long filaments \cite{cornelissen2018cell} ($N_{\rm F}$ = 60; see above). The fibers themselves consist of a conductive core (diameter $d_{\rm F}$ = 26 nm), surrounded by an electrically insulating shell \cite{boschker2021efficient}. The large ratio of length (1~cm) to diameter (26~nm) suggests that the electron transport in the fibers is one-dimensional. Still, there could be multiple conduction channels acting in parallel in a single fiber, as $d_{\rm F}$ far exceeds the typical spacing between charge carrier sites in metalloproteins ($1-2$~nm). 

At present, it is not known how many conduction channels are actually present. Here, we assume that a conductive fiber core consists of multiple parallel molecular conduction paths, each $d_{\rm P} = 2$ nm in diameter. Assuming a hexagonal packing of cylinders, the number of parallel channels in one periplasmic fiber can then be estimated as $N_{\rm C} = 125$. The total number of parallel paths in a given cable bacterium filament thus amounts to $N_{\rm P} = N_{\rm F}\cdot N_{\rm C} =$ 7500. This large number of conduction channels should be regarded as an upper limit of what's physically possible, thus providing a conservative estimate for the electron transfer rate. If less channels are present, then the conductivity of individual channels must be higher, and this in turn also necessitates higher electron transfer rates (see discussion in main text).       

\nolinenumbers
\subsubsection*{Current through a single conduction channel}
\nolinenumbers
Each conduction channel is modelled as a one-dimensional chain of $N_{\rm S}$ equidistant hopping sites, which serve as temporary localization centers of an electron during the hopping process (fig.~S\ref{fig:HoppingSchematic}). The site at position, $i$, along the chain is characterised by the site energy, $U_{\rm i}$. Transitions between sites represent the quantum-mechanical tunneling of electrons, and are characterised by a forward rate, $\Gamma_{\rm i,j}$, and a backward rate, $\Gamma_{\rm j,i}$. The ratio between forward and backward rates is governed by the detailed balance relation \cite{egger1994quantum,Nazarov2009QuantumTI}:
\begin{linenomath}
\begin{equation}
    \Gamma_{\rm i,j}/\Gamma_{\rm j,i} = \text{exp}(\Delta/k_{\rm B} T). 
\end{equation}
\end{linenomath}
Here, $k_{\rm B}$ is the Boltzmann constant, $T$ is temperature, and $\Delta=U_{\rm i}-U_{\rm j}$ is the driving force, i.e., the difference between the site energy of the initial position of the electron, $U_{\rm i}$, and that of its final position, $U_{\rm j}$. In our model, only nearest neighbour hopping is considered ($j = i \pm 1$). Moreover, the hopping sites are assumed to be identical and periodic in structure, with the same center-to-center distance $a$ between them (fig.~S\ref{fig:HoppingSchematic}). As sites are identical, the transition rate is the same for every pair of sites ($\Gamma_{\rm i,i+1}=\Gamma_{\rm R}$; $\Gamma_{\rm i+1,i}=\Gamma_{\rm L}$). Moreover, if we connect the hopping chain to an electrode on either side, and impose a voltage bias  $V$ between the electrodes, the energy difference between two sites is constant and amounts to $\Delta = eV/N_{\rm S}$ (assuming there is no voltage drop over the electrode interface). The current between two consecutive hopping sites along a single conduction path is given by: 
\begin{linenomath}
\begin{equation}
I_{\rm P}= e \left[ \Gamma_{\rm R} p_{\rm i} \left(1-p_{\rm i+1} \right)-\Gamma_{\rm L}p_{\rm i+1}\left(1-p_{\rm i} \right) \right],
\end{equation}
\end{linenomath}
where $p_{\rm i}$ and $p_{\rm {i+1}}$ are the occupation probabilities at the consecutive hopping sites, and $e$ is the elementary charge (i.e., the magnitude of the charge carried by a single electron). This expression embeds the Pauli exclusion principle: an electron can only jump to a charge carrier site when this site is vacant. 

Because the current is not dependent on the pair of sites where it is calculated, only certain combinations of $p_i$ are possible. We make the simplification that our hopping chain is optimally filled for conductance ($p_{\rm i} = 1/2$), using the following justification. Because in the electrode many electron states can contribute to the effective hopping rates at the interface, these interfacial rates can be expected to be far greater than the internal hopping rates, $\Gamma_{\rm R,L}$. The occupation probability of the edge hopping sites, $p_{\rm 1}$ and $p_{\rm N}$, are then expected to follow the Fermi-Dirac distribution in the metal electrodes:
\begin{linenomath}
\begin{equation}
p_{\rm FD} = \frac{1}{1+\text{exp}\left(\frac{\epsilon}{k_{\rm B} T}\right)}.  
\end{equation}
\end{linenomath}
Here, $p_{\rm FD}$ is the occupation probability of an electron state inside the electrode and $\epsilon$ is the energy difference of this electron state with respect to the energy of the electrode. If $\epsilon=0$ for $i={1,N}$ (see fig.~S\ref{fig:HoppingSchematic}), then $p_1=p_N=1/2$ and no occupancy difference is expected to build up in the hopping chain, because all internal hopping rates $\Gamma_{\rm R,L}$ are the same. Therefore,  all sites in the chain can be expected to have the occupancy $p_i = 1/2$. 

Accounting for $N_{\rm P}$ parallel conduction channels, and implementing the detailed balance relation, the total current along a single filament, can be expressed as:
\begin{linenomath}
\begin{equation}
I = N_{\rm P} I_{\rm P} = e \frac{1}{4}N_{\rm P}(\Gamma_{\rm R}-\Gamma_{\rm L}).
\label{eq:BasicCurrent}
\end{equation}
\end{linenomath}
Implementing the detailed balance relation, we obtain:
\begin{linenomath}
\begin{equation}
I = e \frac{1}{4}N_{\rm P} \Gamma_{\rm R} \left(1-\text{exp}(-\Delta/k_{\rm B} T)\right).
\label{eq:StandardCurrent}
\end{equation}
\end{linenomath}
Two separate regimes can be discerned. In the low-bias, high-temperature regime ($ k_{\rm B} T > eV/N_{\rm S})$, the current in Eq.~\ref{eq:StandardCurrent} simplifies to:
\begin{linenomath}
\begin{equation}
I \approx \frac{e}{4}N_{\rm P} \Gamma_{\rm R} \frac{eV}{N_{\rm S} k_{\rm B} T}.
\label{eq:Ihigh}
\end{equation}
\end{linenomath}
Alternatively, in the low-temperature, high-bias regime ($eV/N_{\rm S}  > k_{\rm B} T$), the exponential term in Eq.~\ref{eq:StandardCurrent} vanishes, and the current reduces to:
\begin{linenomath}
\begin{equation}
I \approx e \frac{1}{4}N_{\rm P} \Gamma_{\rm R}.
\label{eq:Ilow}
\end{equation}
\end{linenomath}
The objective is now to find a suitable expression for the transition rate, $\Gamma_{\rm R}$.

\nolinenumbers
\subsubsection*{Transition rate estimates}
\nolinenumbers

The fiber conductivity $\sigma$ can be obtained from the current expression in the low-temperature, high-bias regime, Eq.~\ref{eq:Ihigh}, as:
\begin{linenomath}
\begin{equation}
\sigma = \frac{I}{V}\frac{L}{N_{\rm F}A_F} = \Gamma_{\rm R} \frac{e^2}{k_{\rm B} T}\frac{L N_{\rm C}}{N_{\rm S} \pi d_{\rm F}^2},
\label{eq:ConductivityD}
\end{equation}
\end{linenomath}
As described above, the fiber sheath has $N_{\rm F} = 60$ fibers with a diameter of $d_{\rm F} = 26$ nm, and a cross-sectional area $A_F= \pi d_{\rm F}^2 /4$. Moreover, we estimate that a maximum of $N_{\rm C} = 125$ parallel conduction channels can be integrated in a single fiber (see 'Geometry of the conductive network'). The above expression can be rearranged to provide the transition rate $\Gamma_R$ as a function of the measured conductivity 
\begin{linenomath}
\begin{equation}
\Gamma_{\rm R} = \sigma \frac{k_{\rm B} T}{e^2} \frac{\pi d_{\rm F}^2}{aN_{\rm C}},
\label{eq:ConductivityD}
\end{equation}
\end{linenomath}
In this, $a=L/N_{\rm S}$ represents the mean center-to-center distance between two consecutive sites. 

Conductivities on the order of $\sigma = 100$ S/cm have been estimated for the cable bacterium fibers (see also fig.~S\ref{fig:EXT_FIG_LengthDep}). Moreover, typical center-to-center distances, $a$, between the cofactors in metalloproteins are in the range of 0.5 < $a$ < 2 nm. For a typical cable bacterium filament ($L$ = 1 cm) this hence translates in between 5 and 20 million sites along the hopping chain. Using these numbers, Eq.~\ref{eq:ConductivityD} predicts that the transition rate must range between $\Gamma_R$ = 13 $-$ 54 x 10$^{12}$ s$^{-1}$. This hopping rate exceeds the speed limit of 10$^{13}$ s$^{-1}$ for non-adiabiatic electron transfer \cite{polizzi2012physical}, thus illustrating the theoretical challenge of applying multistep hopping models in cable bacteria (see discussion in main text). 

\nolinenumbers
\subsection*{Transition rate models}
\nolinenumbers
At low temperatures, the energy stored in vibrational modes can stimulate  the hopping process. To account for this effect, several authors have formulated an extended, quantum version of the semi-classical Marcus electron transfer theory. These models explicitly account for vibrational modes that suitably couple to the electron transfer. Two prominent examples are the single effective mode formulation presented by Jortner \cite{Jortner1976TemperatureDA} and the multiple mode formulation developed by Egger \cite{egger1994quantum}. Below, we will integrate these two vibrational model formulations in the multistep hopping chain model introduced above. But first, we introduce the classical Marcus rate formalism as a reference for these more elaborate quantum-based models.

\subsubsection*{High-temperature regime: the Marcus model formulation}

At a sufficiently high temperatures, the thermal energy, $k_{\rm B} T$, sufficiently exceeds the quantum energy, $\hbar\omega$, of the intramolecular vibrations, and all vibrations are then thermally excited. Under these conditions, the transition rate between neighbouring sites is given by the classical Marcus rate expression \cite{taylor2018generalised},
\begin{linenomath}
\begin{equation}
\Gamma_{\rm R} = \frac{2\pi}{\hbar}\frac{V^2_{\rm DA}}{\sqrt{4\pi\lambda k_{\rm B} T}}\text{exp}\left[\frac{-(\lambda - \Delta)^2}{4\lambda k_{\rm B} T}\right].
\label{eq:MarcusRate}
\end{equation}
\end{linenomath}
Here, $\lambda$ is the reorganization energy, $V_{\rm DA}$ is the electronic coupling, and $\Delta= eV/N_{\rm S}$ is the energy difference between adjacent sites introduced above. Upon substitution of this expression into the high-temperature regime current (Eq. \ref{eq:Ihigh}), we obtain the conductance:
\begin{linenomath}
\begin{equation}
G =  \frac{e^2N_{\rm P}}{4 N_{\rm S} k_{\rm B} T}\frac{2\pi}{\hbar}\frac{V^2_{DA}}{\sqrt{4\pi\lambda k_{\rm B} T}}\text{exp}\left[{\frac{-(\lambda-\frac{eV}{N_{\rm S}})^2}{4\lambda k_{\rm B} T}}\right].
\label{eq:MarcusG}
\end{equation}
\end{linenomath}
Because the bacterial filaments investigated are very long ($N_{\rm S}>10^2$), the driving force $\Delta = eV/N_{\rm S}$ in the low-bias regime ($V<0.5$ V) will be < 5 meV, which is substantially smaller than the reorganization energy, $\lambda \approx 270$ meV (Fig.~\ref{fig:4}). Consequently, the exponent in Eq.~\ref{eq:MarcusG} reduces to $e^{-\lambda/4 k_{\rm B} T}$ and the conductance relation simplifies to:
\begin{linenomath}
\begin{equation}
G = g_0 (k_{\rm B} T)^{-3/2}\text{exp}(-\frac{\lambda}{4k_{\rm B}T}).
\label{eq:MarcusGbis}
\end{equation}
\end{linenomath}
The pre-factor $g_0$ is given by:
\begin{linenomath}
\begin{equation}
g_0  =  \frac{e^2N_{\rm P}}{4 N_{\rm S}} \frac{2\pi}{\hbar}\frac{V^2_{\rm DA}}{\sqrt{4\pi\lambda}}.
\label{eq:Marcus_g0}
\end{equation}
\end{linenomath}
Eq. \ref{eq:MarcusGbis} was fitted to the $G(T)$ data to provide values for $\lambda$ and $g_0$ (non-linear least squares fit of $G$ versus $1/T$ as displayed in Fig.~\ref{fig:4}A; values for $\lambda$ and $V_{\rm DA}$ tabulated in table~S\ref{table:NT}).

This fitting procedure assumes that $g_0$ itself is temperature independent. However, it is known that the electronic coupling $V_{\rm DA}$ is weakly temperature dependent \cite{giannini2019quantum}. Consequently, the exact temperature dependence of the pre-exponential factor in Eq. \ref{eq:MarcusGbis} is unknown. As a simplification, one can leave the $T^{-3/2}$ dependence out of the equation, to arrive at the classical Arrhenius relation:
\begin{linenomath}
\begin{equation}
G = g_1 \text{exp}\left(-\frac{U_{\rm A}}{k_{\rm B} T}\right).
\label{eq:Arrhenius}
\end{equation}
\end{linenomath}
Here, the pre-factor $g_1 = g_0 (k_{\rm B} T)^{-3/2}$ is now considered to be a temperature-independent parameter. This basic Arrhenius relation was also fitted to the high temperature conductance data to arrive at values for the activation energy $U_{\rm A}$ (fig.~S\ref{fig:HighT_G_Analysis}; table~S\ref{table:basic}).

\nolinenumbers
\subsubsection*{Low-temperature regime, one effective vibrational mode: the Jortner model formulation}
\nolinenumbers

When the quantum energy of vibrations $\hbar\omega$ becomes smaller than the thermal energy, $k_{\rm B} T$, the Marcus rate expression (Eq.~\ref{eq:MarcusRate}) is no longer valid.  In the Jortner model formulation, vibrational modes are characterised by a single effective mode $\langle \omega \rangle$, which provides a suitable average across all molecular vibrations. The corresponding transition rate becomes \cite{Jortner1976TemperatureDA}:
\begin{linenomath}
\begin{equation}
    \Gamma_{\rm R} = \Gamma_0 \cdot \text{exp}(-S(2\overline{n}_{\rm B}+1))\cdot I_{\rm q} (2S(\overline{n}_{\rm B}(\overline{n}_{\rm B}+1))^{1/2})\cdot (\overline{n}_{\rm B}(\overline{n}_{\rm B}+1))^{q/2}.
\label{eq:Jortner}
\end{equation}
\end{linenomath}
In this expression, $\Gamma_0$ is a temperature-independent pre-factor, $S=\lambda/\hbar\langle\omega\rangle$ is the electron-phonon coupling strength associated with $\langle \omega \rangle$,  $\overline{n}_{\rm B} = 1/(\text{exp}(\hbar\langle\omega\rangle/k_{\rm B} T)-1)$ is the Bose-Einstein distribution for the single effective mode $\langle \omega \rangle$, the exponent $q = \frac{\Delta}{\hbar\langle\omega\rangle}$ is the ratio between the site energy drop between hopping sites and the mean quantum vibrational energy, and $I_{\rm q}$ is the modified Bessel function \cite{Jortner1976TemperatureDA}. One can show that at high temperatures (when $\hbar\langle \omega \rangle \ll k_{\rm B} T$), the Marcus rate expression is recovered. 

The Jortner expression for the transition rate, Eq.~\ref{eq:Jortner}, can be combined with the expression for the current along a single filament, Eq.~\ref{eq:StandardCurrent}, in order to estimate the conductance $G$ as a function of temperature and the electric field strength. The results are given in Fig.~\ref{fig:3}C and D. At high temperatures, the conductance follows an Arrhenius type dependence (Fig.~\ref{fig:3}C), while at low temperatures ($k_{\rm B} T \ll \hbar\langle\omega\rangle$), the conductance becomes temperature-independent (Fig.~\ref{fig:3}D). This calculated $G(T,E)$ response shows a good agreement with the data.      

\nolinenumbers
\subsubsection*{Low-temperature regime, multiple vibrational modes: the Egger model formulation}
\nolinenumbers

In the Egger model formulation\cite{egger1994quantum}, it is assumed that many vibrational modes are coupled to the electron hopping process. The phonon spectral density function $J(\omega)$ describes how much a certain vibrational mode contributes to the reorganization energy\cite{Kell2013OnTS}:
\begin{linenomath}
\begin{equation}
\lambda = \hbar \int_0^\infty \omega {J(\omega)}{\rm d} \omega.
\end{equation}
\end{linenomath}
The Egger model formulation assumes that the phonon spectral density function has an Ohmic form and can be expressed as:
\begin{linenomath}
\begin{equation}
J(\omega)= (\beta + 2) \frac{e^{-\omega/\omega_{\rm c}}}{\omega}.
\end{equation}
\end{linenomath}
Here, $\omega_{\rm c}$ is the characteristic frequency of the spectrum. For the Ohmic spectral density, the reorganization energy is directly proportional to the characteristic frequency, $\omega_{\rm c}$, as stated in the original derivation of the Egger model \cite{egger1994quantum}. :
\begin{linenomath}
\begin{equation}
\lambda = (\beta + 2)\hbar\omega_{\rm c},
\label{eq:hbar_omega_c}
\end{equation}
\end{linenomath}
Since the spectrum starts at zero frequency, some modes remain thermally activated at the lowest temperatures. This explains an important difference between the Jortner and Egger model formulations. Whereas in the Jortner model, the low-bias conductance remains constant at cryogenic temperatures, this is not the case in the Egger model: the low-bias conductance shows a power law behaviour, in which the conductance further decreases as the temperature decreases. Effectively, the modes that remain thermally activated at low temperatures force the low-bias conductance to die out.

When the temperature is much lower than the characteristic energy of vibration ($k_{\rm B} T \ll \hbar\omega_{\rm c}$), the Egger model provides following expression for the transition rate \cite{egger1994quantum}:
\begin{linenomath}
\begin{equation}
\Gamma_{\rm R} = \frac{1}{\gamma(\beta+2)\hbar}\frac{V_{\rm DA}^2}{\hbar\omega_{\rm c}}\left(\frac{2\pi k_{\rm B} T}{\hbar\omega_{\rm c}}\right)^{\beta+1}\left\lvert\gamma\left(1+\beta/2+i\frac{eV}{2\pi N_{\rm S} k_{\rm B} T}\right)\right\rvert^2 \text{exp}\left(\frac{eV}{2N_{\rm S} k_{\rm B} T}\right).  
\label{eq:EggerRate}
\end{equation}
\end{linenomath}
Here, $\gamma$ is the complex gamma function and the exponent $\beta$ provides a measure for the electron-phonon coupling strength. If we combine the expression for the current along a single filament, Eq.~\ref{eq:StandardCurrent}, with the above expression for the transition rate, Eq.~\ref{eq:EggerRate} , we obtain:
\begin{linenomath}
\begin{equation}
I=A_0 T^{\beta+1} \text{sinh}\left(\frac{eV}{2\pi N_{\rm S} k_{\rm B} T}\right)\left\lvert\gamma\left(1+\beta/2+i\frac{eV}{2\pi N_{\rm S} k_{\rm B} T}\right)\right\rvert^2.
\label{eq:USC}
\end{equation}
\end{linenomath}
The pre-factor $A_0$ combines all temperature-independent parameters. Note, that the Egger model centrally depends on the assumption that the phonon spectral density function has an Ohmic form. However, at present, the details of electron-phonon coupling in the conductive fibers of cable bacteria are unknown, and so, alternative shapes phonon spectral density function could be relevant \cite{Kell2013OnTS}.

\nolinenumbers
\subsubsection*{Universal scaling curve}
\nolinenumbers

Closer inspection of Eq.~\ref{eq:USC} reveals that the $I(V)$ data can be replotted as a single universal scaling curve, provided one normalizes the bias voltage as $eV / (2\pi k_{\rm B} T)$ (the horizontal axis) and rescales the current as $I/T^{\beta + 1}$ (vertical axis). If the Egger model applies, then the $I(V)$ data collected at different temperatures should fall onto a single curve, only depending on $A_0$, $\beta$, and the number of hopping sites $N_{\rm S}$. These universal scaling plots are shown in fig.~S\ref{fig:USC_All}. In the low-bias, high-temperature regime ($eV/N_{\rm S} < k_{\rm B} T$), the approximation $\text{sinh}(x)\approx x$ can be made and then $G = I/V \propto T^{\beta}$. Accordingly, one expects the conductance to follow a power law dependence on temperature, which is also seen in the data (Fig.~\ref{fig:4}C). Oppositely, in the high-voltage, low-temperature regime ($eV/N_{\rm S} > k_{\rm B} T$), the transition  rate is given by \cite{egger1994quantum}:

\begin{linenomath}
\begin{equation}
\Gamma_{\rm R} = \frac{2\pi}{\hbar}\frac{V^2_{DA}}{\gamma(\beta+2)\hbar\omega_{\rm c}}\left(\frac{eV}{N_{\rm S} \hbar \omega_{\rm c}}\right)^{\beta+1} \text{exp}\left( -\frac{eV}{N_{\rm S}\hbar\omega_{\rm c}} \right).
\label{eq:LowTRate}
\end{equation}
\end{linenomath}
If we combine this expression with that for the current in the high-bias, low-temperature regime (Eq.~\ref{eq:Ilow}), the conductance becomes: 
\begin{linenomath}
\begin{equation}
G = \frac{e^2}{4}\frac{N_{\rm P}}{N_{\rm S}}  \frac{2\pi}{\hbar}\frac{V^2_{DA}}{\gamma(\beta+2)(\hbar\omega_{\rm c})^2}\left(\frac{eV}{N_{\rm S} \hbar \omega_{\rm c}}\right)^{\beta}. 
\label{eq:LowG}
\end{equation}
\end{linenomath}
In this equation, we have dropped the exponential decay factor in  Eq.~\ref{eq:LowTRate}, as one can prove that for small $x= eV / (N_{\rm S}\hbar\omega_{\rm c})$, this term is very close to one. The overall result is that the conductance depends on the bias voltage via a power law: $G\propto V^{\beta}$. This relation is indeed seen in the data (Fig.~\ref{fig:4}D). 

\nolinenumbers
\subsubsection*{The distance between hopping sites}
\nolinenumbers

Each conduction channel is modelled as a one-dimensional chain of $N_{\rm S}$ equidistant hopping sites with the same center-to-center distance $a=L/N_{\rm S}$ between them (fig.~S\ref{fig:HoppingSchematic}). We can estimate the number of hopping sites $N_{\rm S}$ from our data in two independent ways. In a first approach, we can extract $N_{\rm S}$ directly from the universal scaling curve (fig.~S\ref{fig:USC_All}). The scaling curves show a pronounced inflection point at the transition from the low-bias, high-temperature regime towards the high-bias, low-temperature regime. This transition occurs at a bias 
\begin{linenomath}
\begin{equation}
eV \approx 2\pi N_{\rm S} k_{\rm B} T
\label{eq:NSfrom}
\end{equation}
\end{linenomath}
As a result, the value of $eV /2\pi k_{\rm B} T$ at the inflection point of the scaling curve provides a direct estimate of the apparent number of hopping steps $N_{\rm S}$. In fig.~S\ref{fig:USC_All}, panel (O), this transition point is indicated by the dotted line. In table~S\ref{table:NT}, the resulting values of $N_{\rm S,USC}$ are listed for each of the 16 segments investigated. The corresponding estimate for the center-to-center distance $a_{\rm USC} = L/N_{\rm S,USC}$ ranges between 33 and 102 nm.

In the second approach, a value for $N_{\rm S}$ is numerically calculated from other parameters available. To this end, we first note that for low temperatures and high voltages ($eV/N_{\rm S} \gg k_{\rm B} T$), the conductance scales as $G = B_0 V^{\beta}$, where $B_0$ is the pre-factor in the power law fit. Using Eq.~\ref{eq:LowG}, this pre-factor can be explicitly written as: 
\begin{linenomath}
\begin{equation}
B_0 = \frac{e^2}{4}\frac{N_{\rm P}}{N_{\rm S}}  \frac{2\pi}{\hbar}\frac{V^2_{DA}}{\gamma(\beta+2)(\hbar\omega_{\rm c})^2}\left(\frac{e}{N_{\rm S} \hbar \omega_{\rm c}}\right)^{\beta}. 
\label{eq:B0}
\end{equation}
\end{linenomath}
Rearrangement provides an expression for the number of hopping sites:  
\begin{linenomath}
\begin{equation}
    N_{\rm S} = \frac{e}{\hbar\omega_{\rm c}}\left[\frac{eN_{\rm P}}{4} \frac{2\pi}{\hbar} \frac{V^2_{\rm DA}}{\hbar\omega_{\rm c} \gamma(\beta+2)}\frac{1}{B_0}\right]^{1/(\beta+1)}.
    \label{eq:NS_BASIC}
\end{equation}
\end{linenomath}
This expression requires an estimate for the electronic coupling $V_{\rm DA}$. To arrive at this, we note that the pre-factor $g_0$ (as derived from the fitting of the Marcus rate expression, Eq.~\ref{eq:Marcus_g0}) can be transformed into an expression for the electronic coupling:
\begin{linenomath}
\begin{equation}
V_{\rm DA}^2  =  \left[g_0 \frac{\hbar}{2\pi} \frac{4N_{\rm S}}{e^2 N_{\rm P}} \sqrt{4\pi \lambda} \right].
\label{eq:Marcus_VDA}
\end{equation}
\end{linenomath}
Substitution of Eq.~\ref{eq:Marcus_VDA} into Eq.~\ref{eq:NS_BASIC} finally gives: 
\begin{linenomath}
\begin{equation}
    N_{\rm S} = \frac{e}{\hbar\omega_c}\left[\frac{\sqrt{4\pi \lambda}}{\gamma(\beta+2) (\hbar\omega_{\rm c})^2}\frac{g_0}{B_0}\right]^{1/\beta}.
    \label{eq:NS_FINAL}
\end{equation}
\end{linenomath}
From this, $N_{\rm S}$ can be calculated provided that we supply parameters that are fitted in the low-temperature regime (the pre-factor $B_0$, the coupling exponent $\beta$, the characteristic frequency $\hbar\omega_{\rm c}$) as well as parameters that are determined in the high-temperature regime (pre-factor $g_0$, reorganization energy $\lambda$).  

The above derivation essentially assumes that the electronic coupling is independent of temperature (the high-temperature value is substituted into the low-temperature expression). It is, however, known that $V_{\rm DA}$ is enhanced by thermal fluctuations. So in reality, the value of $V_{\rm DA}$ could  be smaller at lower temperatures, and as a consequence, the value of $N_{\rm S}$ will be overestimated by Eq.~\ref{eq:NS_FINAL}. This could explain the differences obtained for $N_{\rm S}$ by the two separate estimation procedures (see fig.~S\ref{fig:NL}). However, in high-mobility organic semiconductors, this temperature effect changes the $V_{\rm DA}$ value by less than one order of magnitude \cite{giannini2019quantum}. Because of the power law dependence $1/\beta$ in Eq.~\ref{eq:NS_FINAL}, such a tenfold difference in electronic coupling only causes a two-fold difference in the $N_{\rm S}$ estimate.

The resulting values of $N_{\rm S,NUM}$ are listed in table~S\ref{table:NT} for all 16 segments investigated. The associated center-to-center distance $a_{\rm NUM} = L/N_{\rm S,NUM}$ ranges between 10 and 36 nm, which is a factor of 3 lower than the corresponding estimate based on the universal scaling curve. fig.~S\ref{fig:NL} shows that the number of hopping sites linearly scales with the segment length. This trend is seen for both approaches.

\end{appendices}

\end{document}